\documentclass[aps,prb,twocolumn,superscriptaddress]{revtex4-1}
\usepackage[english]{babel}
\usepackage{amsmath,amsthm}
\usepackage{amsfonts}
\usepackage[pdfborder={0 0 0}, colorlinks=true, urlcolor=blue, linkcolor=blue, citecolor=blue]{hyperref}
\usepackage{color}
\usepackage{graphicx}
\usepackage{epstopdf}
\usepackage{float}
\usepackage{subfigure}
\usepackage{ulem}

\theoremstyle{definition}

\theoremstyle{remark}

\begin{document}

\title{Stability of vortices in exciton-polariton condensates with spin-orbital-angular-momentum coupling}

\author{Xin-Xin Yang}
\affiliation{Department of Physics and Beijing Key Laboratory of Opto-electronic Functional Materials and Micro-nano Devices, Renmin University of China, Beijing 100872, China}
\affiliation{Key Laboratory of Quantum State Construction and Manipulation (Ministry of Education), Renmin University of China, Beijing 100872,China}
\author{Wei Zhang}
\thanks{wzhangl@ruc.edu.cn}
\affiliation{Department of Physics and Beijing Key Laboratory of Opto-electronic Functional Materials and Micro-nano Devices, Renmin University of China, Beijing 100872, China}
\affiliation{Key Laboratory of Quantum State Construction and Manipulation (Ministry of Education), Renmin University of China, Beijing 100872,China}
\affiliation{Beijing Academy of Quantum Information Sciences, Beijing 100872, China}
\author{Zhen-Xia Niu}
\thanks{niuzhx@zjnu.edu.cn}
\affiliation{Department of Physics, Zhejiang Normal University, Jinhua 321004, China}
\date{\today }

\begin{abstract}

The existence and dynamics of stable quantized vortices is an important subject of quantum many-body physics. Spin-orbital-angular-momentum coupling (SOAMC), a special type of spin-orbit coupling, has been experimentally achieved to create vortices in atomic Bose-Einstein condensates (BEC). 
Here, we generalize the concept of SOAMC to a two-component polariton BEC and analyze the emergence and configuration of vortices under a finite-size circular pumping beam.
We find that the regular configuration of vortex lattices induced by a finite-size circular pump is significantly distorted by the spatially dependent Raman coupling of SOAMC, even in the presence of a repulsive polariton interaction which can assist the forming of stable vortex configuration. Meanwhile, a pair of vortices induced by SOAMC located at the center of polariton cloud remains stable. When the Raman coupling is sufficiently strong and interaction is weak, the vortices spiraling in from the edge of polariton cloud will disrupt the polariton BEC.

\end{abstract}

\maketitle

\section{INTRODUCTION}
Quantized vortices, as a type of topological defects, have been observed and extensively studied in many physical systems such as superconductors~\cite{ESSMANN1967526}, superfluid liquid helium~\cite{Williams1974,Yarmchuk1979}, cold atoms~\cite{Madison2000,abo2001observation}, photon fields~\cite{Zhen2014,Zijian2018}, and exciton-polariton Bose-Einstein condensates (BEC)~\cite{lagoudakis2008,Lagoudakis2009}. In many of these notable examples, quantized phase winding and/or rotational superflow have been witnessed as a key evidence of superfluidity/superconductivity. Formed by strongly coupled cavity photon and quantum-well exciton in a semiconductor microcavity, exciton-polariton BEC, or polariton BEC for short, has the advantage of high transition temperature and versatile optical control/probe owing to its photonic component~\cite{QihuaXiong2021,Real2021}. 
It is also conceptually interesting as polariton BEC is a non-equilibrium quantum system with macroscopic coherence, where the polariton continuously decays and simultaneously replenishes via stimulated scattering of reservoir-exciton excited by the optical pump~\cite{Deng2010,Levinsen2019}. The non-equilibrium nature and photonic component of polariton BEC provide a unique opportunity for investigating the dynamics of vortices~\cite{lagoudakis2008,Panico2021,Ardizzone2022}, which can be created by various experimental techniques such as resonant excitation by Laguerre-Gaussian (LG) beams~\cite{boulier2015,kwon2019,Choi2022}, optical parametric oscillation~\cite{sanvitto2010}, and pumping with a circular excitation beam~\cite{Boulier2016,Hu2020}. 

Another stark characteristic of polariton BEC is the existence of a pseudo-spin degree of freedom inherited from the angular momentum coupling between exciton and photon components. Heavy-hole excitons with different magnetic quantum numbers can couple to photons with different circular polarizations, leading to a splitting of energy between the two composite exciton-polariton states, which can be looked at as two pseudo-spin states with $S=1/2$~\cite{Panzarini1999}. This so-called transverse-electric-transverse-magnetic (TE-TM) splitting depends on the in-plane wave vector~\cite{Kavokin2005,Tassone1992}, and takes the role of Zeeman energy in the pseudo-spin space. Notable effects are discussed therein, including the optical spin-Hall effect~\cite{Leyder2007,Kammann2012,Kavokin2005}, spin vortices~\cite{Dufferwiel2015}, magnetic-monopole-like half solitons~\cite{hivet2012half}, half-quantum circulation~\cite{Gang2015}, ferromagnetic phase transition~\cite{Ohadi2015}, and transition from a highly coherent to a super-thermal state~\cite{Baryshev2022}. 

Recently, one special type of spin-orbit coupling, known as spin-orbital-angular-momentum coupling (SOAMC), has been achieved in quantum gases of neutral atoms by using two copropagating LG laser beams to couple different atomic hyperfine states via a two-photon Raman process~\cite{Lin2018,Jiang2019}. Exotic quantum phases have been proposed and investigated for bosonic and fermionic systems, including the stripe phase, vortex-antivortex-pair phase, and half-skyrmion phase~\cite{Bidasyuk2022,Jian2021,Hui2020,Chen2020}. 

In this work, we propose a scheme to implement SOAMC in two-component polariton BEC in a two-dimensional heterostructure hosting two photonic modes, and study the steady-state solutions and vortex dynamics under a finite-size circular pumping beam. We adopt an open-dissipation Gross-Pitaevskii theory to describe the dynamical evolution of the polariton BEC and examine the stability of steady-state solutions under the balance of gain and loss. 
We find that the presence of SOAMC can stabilize a pair of vortices located at the center of the polariton clouds in the stable state. Furthermore, by analyzing the density and phase distributions in the dynamical evolution, we investigate the effects of Raman coupling of SOAMC and interaction on vortex lattices in the polariton BEC subjected to a finite-size circular pump. We find that while the spatially inhomogeneous Raman coupling tends to break the vortex lattice, a repulsive interaction helps to build a stable edge of polariton cloud and thus hold the pattern of vortices.

The remainder of this paper is organized as follows. In Sec.~\ref{sec:Model}, we propose a scheme to realize SOAMC in polariton BEC and derive a dimensionless form of the open-dissipative Gross-Pitaevskii equation from a single-particle Hamiltonian. In Sec.~\ref{sec:theory}, we analyze the steady-state solutions under adiabatic approximation. The dynamical evolutions with different Raman coupling strengths and interactions are simulated in Sec.~\ref{sec:numerical}. Finally, we summarize in Sec.~\ref{sec:summary}.

\section{Polariton BEC with SOAMC}
\label{sec:Model}

\begin{figure}[tbp]
	\centering 
	\includegraphics[width=0.48\textwidth,height=0.25\textheight]{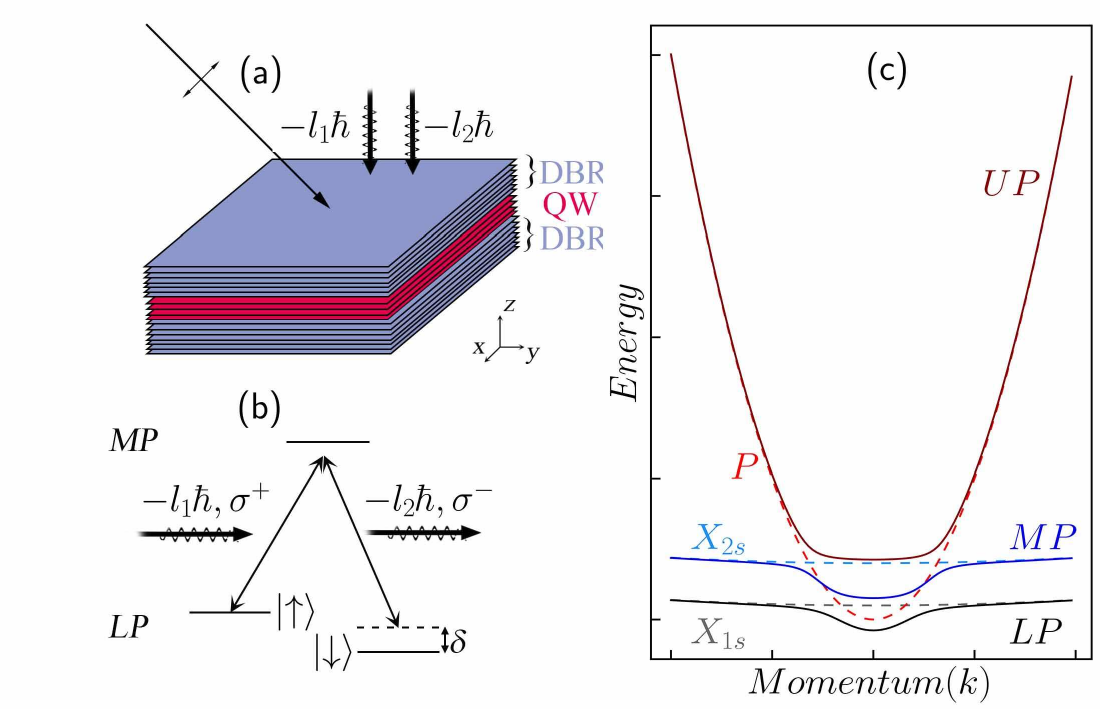}
	\caption{(Color online) A schematic of the exciton-polariton BEC with SOAMC. (a) The QWs are sandwiched by two DBRs. The pumping laser incidents in the direction of an angle with the $z$-axis, and the LG beams are applied along the $z$-axis.
	(b) Schematic illustration of the Raman process between the two LP modes. (c) Spectrum of exciton-polariton dispersion. Exciton modes $X_{1s}$ and $X_{2s}$ are represented by grey and blue dashed lines, respectively, while photon modes $P$ are represented by red dashed line. The three branches of quasiparticle mode, such as LP, MP, and UP, are depicted by black, blue, and red solid lines, respectively.} 
	\label{fig0}
\end{figure}

We propose to realize polariton BEC with SOAMC in a two-dimensional heterostructure consisting of CdTe quantum well (QW) layers sandwiched by two distributed Bragg reflectors (DBRs) as shown in Fig.~\ref{fig0}(a). QW excitons, which are bound states of electron-hole pairs, can be excited by an external pumping laser, and bind with the photons of the semiconductor microcavity by the strong exciton-photon coupling to create quasiparticles of exciton-polariton pairs. In comparison to GaAs-based microcavities, the CdTe-based system features two heavy-hole exciton modes [$X_{1s}$ and $X_{2s}$ shown in Fig.~\ref{fig0}(c)] with much higher binding energy ($\sim 25$ MeV) when the temperature is kept at $4.5$ K. By coupling to the cavity photon mode, the two exciton modes lead to three exciton-polariton branches, including the lower polaritons (LP), the middle polaritons (MP), and the upper polaritons (UP), as illustrated in Fig.~\ref{fig0}(c). At zero momentum $k=0$, the minimal Rabi splitting is around $13$ MeV between the LP and MP branches, and around $3.5$ MeV between the MP and UP branches~\cite{Bruchhausen2008}. 

By applying a linearly polarized non-resonant pumping laser with a tilt angle from the normal-axis ($z$-axis)~\cite{Kavokin2005,Leyder2007}, heavy-holes of the $X_{1s}$ mode with different magnetic quantum numbers can absorb either left or right circularly polarized photons and split into TE and TM modes~\cite{Yang2022}. The two modes are considered as two pseudo-spin states with $S=1/2$. The effective Hamiltonian associated with the TE and TM modes takes the form $\hbar\mathbf{\sigma}\cdot\mathbf{\Omega}_k$ in the pseudo-spin basis\cite{Glazov2015,Maialle1993}, where $\mathbf{\sigma} = (\sigma_x,\sigma_y)$ is the Pauli vector, and $\mathbf{\Omega}_k$ is the effective pseudo-spin precession frequency given by $\mathbf{\Omega}_k=(\Delta_T/\hbar)(k_x^2-k_y^2,2k_xk_y)$ with the TE-TM coupling strength $\Delta_T$. 
Notice that the TE-TM splitting energy is dependent on the incident momentum of the pumping beam and can be tuned precisely by controlling the incident angle~\cite{Baxter1997,Panzarini1999,Kavokin2005}, and it can even be tuned to zero~\cite{Levrat2010,Pinsker2014,Li2015,Yulin2016}.
Furthermore, since the actual angular momentum difference between the two modes is two quanta~\cite{Yang2022}, the two states can be coupled by a Raman process with a pair of copropagating longitudinal optical LG beams with left and right circular polarizations~\cite{kwon2019,Jiang2019,Choi2022}. The Raman scheme is mediated by the upper MP branch, as schematically depicted in Fig.~\ref{fig0}(b). Since the two LG beams also carry orbital angular momenta (OAM) $l_1 \hbar$ and $l_2 \hbar$, the Raman transition will cause a change of angular momentum of the polariton motional degree of freedom in the amount of $l \equiv (l_1 - l_2)/2$. By adiabatically eliminating the MP state, 
as explained in detail in the Supplement material, we can obtain an effective model of a single polariton, whose OAM in coordinate space and spin in pseudo-spin space (with the basis $\{| {\rm TE} \rangle,| {\rm TM} \rangle\}$) are coupled
\begin{eqnarray}
	H_0 =  \left[\begin{matrix}
		-\frac{\hbar^2}{2M}\nabla^2_r - \frac{\hbar^2}{2M r^2}\partial_{\theta}^2 + \frac{\delta}{2}&\Omega_R(r)e^{-i2l\theta}\\
		\Omega_R(r)e^{i2l\theta}&-\frac{\hbar^2}{2M}\nabla^2_r - \frac{\hbar^2}{2Mr^2}\partial_{\theta}- \frac{\delta}{2}\\
	\end{matrix}\right].
	\nonumber \\
\end{eqnarray}
Considering that polaritons formed by cavity photons and two-dimensional (2D) semiconductor excitons have a longer lifetime, only the in-plane motion of polariton BEC is discussed here~\cite{RevModPhys.85.299}.
The 2D vector $\mathbf{r}=(r, \theta)$ is defined in the polar coordinate, $\nabla^2_{r}=r^{-1}\partial_{r}r\partial_{r}$, 
and $M\sim10^{-4}m_{e}$ is the effective mass of polariton combined with QW exciton and microcavity photon with $m_{e}$ the mass of free electron.
Besides, $\Omega_R(r)$ and $\delta$ denote the Rabi frequency and two-photon detuning of the Raman process, respectively. Notice that by writing down this single particle Hamiltonian, we do not consider the pumping and decay of polariton, which will be included later. Then, we apply a unitary transformation $\tilde{H}_0 = U^{-1}H_0U$ with $U={\rm diag}(e^{-il\theta}, e^{il\theta})$ to eliminate the phase of the off-diagonal elements~\cite{Lin2018,Jiang2019,Chen2020}, and obtain
\begin{eqnarray}
	\tilde{H}_0 =  \left[\begin{matrix}
		-\frac{\hbar^2}{2M}\nabla^2_r + \frac{(L_z-\hbar l)^2}{{2M r^2}} + \frac{\delta}{2} & \Omega_R(r)\\
		\Omega_R(r)&-\frac{\hbar^2}{2M}\nabla^2_r + \frac{(L_z+\hbar l)^2}{2Mr^2} - \frac{\delta}{2}\\
	\end{matrix}\right].
	\nonumber \\
\end{eqnarray}
Here, $L_{z}=-i\hbar\partial_{\theta}$ denotes the angular momentum operator of polariton along the $z$-axis, which couples with spin via the SOAMC term $-\hbar l L_{z} \otimes \sigma_z/Mr^{2}$. 
Notice that under the unitary transformation, the basis $\{|\psi_1\rangle, |\psi_2\rangle \}$ of $\tilde{H}_0$ essentially represents the rotations of TE and TM modes and defined as  \(|\psi_1\rangle = e^{il\theta}|{\rm TE}\rangle \) and \(|\psi_2\rangle = e^{-il\theta}|{\rm TM}\rangle\).

In atomic BEC, the two-photon Raman process is extensively utilized to investigate vortex dynamics. It can be employed  to generate vortices\cite{Dum1998}, transfer photon angular momentum to the BEC\cite{Marzlin1997}, and store vortex lattices\cite{Dutton2004}.
Moreover, some novel vortex geometries can be generated by transferring angular momentum through LG beams\cite{Maucher2018}.
Here, to investigate the dynamics of a two-component polariton BEC with SOAMC subjected to pumping and decay, we employ the widely accepted open-dissipative Gross-Pitaevskii equations (ODGPEs) approach, which has been applied to describe the dynamical synchronization~\cite{Li2015}, elementary excitations~\cite{Liang2017}, dynamics of vortex~\cite{Borgh2010} and dark soliton train~\cite{Pinsker2014} of two-component polariton BEC.
The order parameters of the two components $|\psi_1\rangle$ and $|\psi_2\rangle$ are represented by the time-dependent wave functions $\Psi_{1}(\mathbf{r},t)$ and $\Psi_{2}(\mathbf{r},t)$ on the mean-field level.
Here, we assume the pumping and decay rates of TE and TM modes are identical, hence can be simply applied to the rotated modes.
Furthermore, we consider that the exciton reservoir relaxes quickly enough for both components to ensure that the stimulated scattering to each component is unaffected~\cite{Yulin2016}. Under this adiabatic condition, a mean-field approach is sufficient to investigate the dynamics of the polariton BEC~\cite{Sturm2015}.

The wave functions $\Psi_{1,2}(\mathbf{r},t)$ and the density $n_e(\mathbf{r},t)$ of exciton reservoir satisfy a coupled set of ODGPEs,
 \begin{eqnarray}
 	\label{eqn:GPE1}
	i\hbar\frac{\partial\Psi_{1}}{\partial t}&=&\left[-\frac{\hbar^2}{2M}\nabla^2_{r}
	+V_{1}(\mathbf{r},t)+\frac{(L_{z}-\hbar l)^{2}}{2Mr^{2}}+\frac{\delta}{2}\right]\Psi_{1} \nonumber\\
	&+&\Omega_{R}(r)\Psi_{2}+\frac{i\hbar}{2}\left(R_{e}n_{e}-\gamma_{p}\right)\Psi_{1}, \\
	\
	\label{eqn:GPE2}
	i\hbar\frac{\partial\Psi_{2}}{\partial t}&=&\left[-\frac{\hbar^2}{2M}\nabla^2_{r}
	+V_{2}(\mathbf{r},t)+\frac{(L_{z}+\hbar l)^{2}}{2Mr^{2}}-\frac{\delta}{2}
	\right]\Psi_{2} \nonumber\\
	&+&\Omega_{R}(r)\Psi_{1}+\frac{i\hbar}{2}\left(R_{e}n_{e}-\gamma_{p}\right)\Psi_{2},\\
	\
	\label{eqn:GPE3}
	\frac{\partial n_e}{\partial t}&=&P(r)-\left[\gamma_e +R_{e}(|\Psi_{1}|^2+|\Psi_{2}|^2)\right]n_e.	
\end{eqnarray}
Here, $V_{j=1,2}(\mathbf{r},t)=\frac{1}{2}M\omega_{\perp}^{2}r^{2}+g|\Psi_{j}|^2+g_{12}|\Psi_{3-j}|^2+g_{e}n_{e}$ denotes an induced effective potential from the mean-field shift caused by the intra- ($g_{1}=g_{2}=g$) and inter-component ($g_{12}$) interactions, the polariton-exciton interaction $g_e$, and the external harmonic trapping potential with oscillator frequency $\omega_{\perp}$~\cite{Science2007}. The polariton BEC and exciton reservoir are lossy with decay rates $\gamma_p$ and $\gamma_e$, respectively. Furthermore, the exciton reservoir is excited by a non-resonant circular pumping beam $P(r)=\eta P_{th}\Theta(R_p-r)$~\cite{Keeling2008}, where ${\Theta}(r)$ is the unit step function, $R_p$ is the cutoff radius, and $\eta$ is a dimensionless factor. The threshold pumping strength $P_{th}\equiv\gamma_{p}\gamma_e/R_e$ corresponds to the exact balance of amplification and loss of exciton-reservoir density. $R_e$ is the stimulated scattering rate from exciton reservoir to polariton BEC. In the following, we consider a representative example where the OAM of LG beams are $l_1=-2$ and $l_2=0$, as used in some experiments~\cite{kwon2019,Jiang2019,Choi2022}. The spatial dependent Raman coupling $\Omega_{R}(r)=\tilde{\Omega}_{R}(r/w)^{|l_{1}|+|l_{2}|}e^{-2r^{2}/w^{2}}$ is characterized by the coupling strength $\tilde{\Omega}_{R}$ and the waist $w$ of the Raman beam. We also focus on the special case of $\delta=0$, which is both experimental feasible and physically insightful since the SOAMC effect can be revealed more transparently without complications induced by the Zeeman field.

To derive a dimensionless form, we take the length unit $a=\sqrt{\hbar/M\omega_{\perp}}$ and time unit $\tau=2/\omega_{\perp}$ to rewrite Eqs.~(\ref{eqn:GPE1})-(\ref{eqn:GPE3}) as
\begin{eqnarray}
	i\frac{\partial\Psi_{1}}{\partial t}&=&\left[-\nabla^2_{r}+V_{1}(\mathbf{r},t)+\frac{(L^{\prime}_{z}-l)^2}{r^2}\right]\Psi_{1} \nonumber\\
	&+&\Omega_{R}(r)\Psi_{2}+\frac{i}{2}(R^{\prime}_e n_{e}-\gamma_{p}^{\prime})\Psi_{1},
	\label{eqn:DGPE1} \\
	\
	i\frac{\partial\Psi_{2}}{\partial t}&=&\left[-\nabla^2_{r}+V_{2}(\mathbf{r},t)+\frac{(L^{\prime}_{z}+l)^2}{r^2}\right]\Psi_{2} \nonumber\\
	&+&\Omega_{R}(r)\Psi_{1}+\frac{i}{2}(R^{\prime}_e n_{e}-\gamma_{p}^{\prime})\Psi_{2}, 
	\label{eqn:DGPE2}\\
	\
	\frac{\partial n_e}{\partial t}&=&P(r)-\left[\gamma_e^{\prime} +R^{\prime}_e(|\Psi_{1}|^2+|\Psi_{2}|^2)\right]n_e.
	\label{eqn:DGPE3}
\end{eqnarray}
Notice that in the expressions above, coordinate $\mathbf{r}$, time $t$, wave function $\Psi_{j}(\mathbf{r},t) (j=1,2)$, effective potential $V_{j}(\mathbf{r},t)$, density $n_e(\mathbf{r},t)$ of exciton reservoir, pumping rate $P(r)$, and Raman coupling $\Omega_{R}(r)$ are all replaced by their dimensionless counterparts, while other parameters are defined as $g^{\prime}=2g/\hbar\omega_{\perp}a^2$, $g^{\prime}_{12}=2g_{12}/\hbar\omega_{\perp}a^2$, 
$g^{\prime}_e=2g_e/\hbar\omega_{\perp}a^2$, $L^{\prime}_{z}=-i\partial_{\theta}$, $\tilde{\Omega}^{\prime}_{R}=2\tilde{\Omega}_{R}/\hbar\omega_{\perp}$, $w^{\prime}=w/a$,  $R^{\prime}_e=2R_e/\omega_{\perp}a^2$, 
$\gamma^{\prime}_p=2\gamma_p/\omega_{\perp}$, $\gamma^{\prime}_e=2\gamma_e/\omega_{\perp}$. 
In realistic experiments~\cite{PhysRevLett.131.136901,Georgios2011}, typical parameters of polariton BEC can be taken as $g\sim1\mu \text{eV}\mu \text{m}^{2}$,  $g_e = 2g$, $ R_e = 0.01\text{ps}^{-1}\mu m^{2}$, $\gamma_p = 0.4\text{ps}^{-1}$, and $\gamma_e = 0.6\text{ps}^{-1}$.

\section{Steady-state solutions}
\label{sec:theory}

\begin{figure*}
	\centering
	\includegraphics[width=1\textwidth]{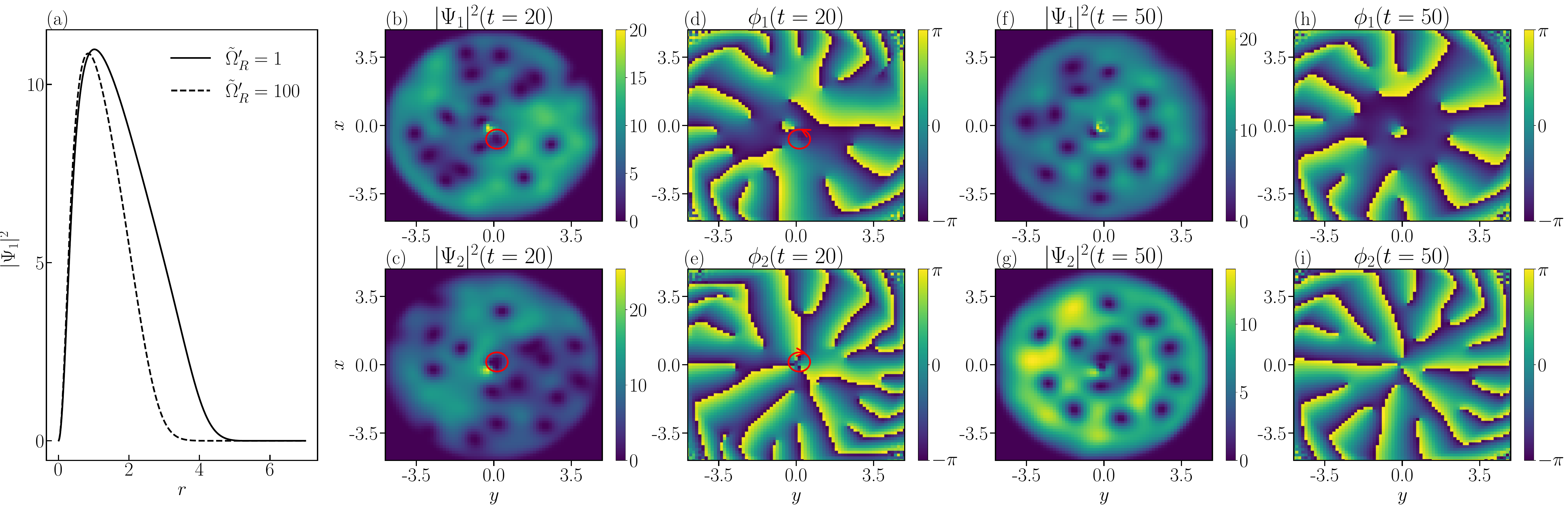}
	\caption{(Color online) (a): The azimuthal symmetric steady-state solutions $n = n_1 = n_2$ of  Eqs.~(\ref{eqn:stead1})-(\ref{eqn:stead2}). (b)-(i): The dynamical evolution of two-component polariton BEC are obtained by numerically solving Eqs.~(\ref{eqn:DGPE1})-(\ref{eqn:DGPE3}) with a Thomas-Fermi-type initial distribution ~(\ref{initial_TF}), 
    (b)-(c) and (f)-(g): the density distributions $|\Psi_{1,2}|^2$, (d)-(e) and (h)-(i): the corresponding phase distributions $\phi_{1,2}$. Other parameters are $\Omega_{v}=1$, $\tilde{\Omega}^{\prime}_R=1$, and $g^{\prime} = 1$.}
	\label{fig1}
\end{figure*}

We consider the steady-state solutions for polariton BEC with SOAMC under a homogeneous pumping and decay. By the definition of a steady state, the net gain is vanishing, such that the density of the exciton reservoir is kept at $n^s_{e}(\mathbf{r})=\gamma^{\prime}_{p}/R^{\prime}$, and the total density is $|\Psi^s_{1}(\mathbf{r})|^{2}+|\Psi^s_{2}(\mathbf{r})|^{2}=(\eta-1)P_{th}/\gamma^{\prime}_p$ when $\eta>1$. In the absence of SOAMC, with $\Omega_{R}(r)=0$ and $l=0$ in Eqs.~(\ref{eqn:DGPE1})-(\ref{eqn:DGPE2}), particles will not exchange between the two pseudo-spin states. The steady-state solution of the ODGPEs without pump and decay is well approximated by the stationary Thomas-Fermi solution, where the velocity (phase gradient) of polariton BEC is zero. While the pumping laser is a finite-size circular beam, the approximate Thomas-Fermi solution will become unstable in a harmonic trap.
Especially, when the radius of the pumping beam is beyond the size of polariton cloud, the net gain at the edge of polariton introduces an instability. Vortices can spiral in from the edge into polariton cloud and spontaneously form a vortex lattice. Thus, a vortex solution can be achieved in polariton BEC pumped by the finite-size circular pumping beam, similar to what is observed in rotating BEC~\cite{Keeling2008,Borgh2012}.

In the presence of SOAMC, we first discuss the scenario of an infinitely large circular pump with $R_p\rightarrow\infty$. The steady-state wave functions $\Psi_{1,2}(\mathbf{r})=\sqrt{n_{1,2}(\mathbf{r})}e^{i\phi_{1,2}(\mathbf{r})}$ are assumed to describe both the density $n_{1,2}(\mathbf{r})$ and phase $\phi_{1,2}(\mathbf{r})$ distributions. Upon substituting the wave functions into Eqs.~(\ref{eqn:DGPE1})-(\ref{eqn:DGPE2}) and focusing on the imaginary parts, we gain the following equations for phase gradients under the Thomas-Fermi approximation,
\begin{eqnarray}
	\label{eqn:phase_part11}	
	\nabla\cdot\left[ n_{1}\left( \nabla\phi_{1}-\frac{\mathbf{l}}{r^2}\times\mathbf{r}\right) \right]&=&-\Omega_{R}(r)\sqrt{n_{1}n_{2}}\sin (\phi_{1}-\phi_{2}), \nonumber\\
	\\
	\nabla\cdot\left[ n_{2}\left( \nabla\phi_{2}+\frac{\mathbf{l}}{r^2}\times\mathbf{r}\right) \right]&=&\Omega_{R}(r)\sqrt{n_{1}n_{2}}\sin (\phi_{1}-\phi_{2})\nonumber,	
\end{eqnarray}
where $\mathbf{l}=l\hat{z}$, and $\hat{z}$ is the unit vector along the $z$-axis. 

If the Raman coupling $\Omega_R(r)$ is negligible, the polarization of two-component polariton BEC can be safely ignored to give $n_{1}=n_{2}=n$. By adding the two equalities of Eq.~(\ref{eqn:phase_part11}), we then obtain
\begin{eqnarray}	
	\label{eqn:phase_part12}
	\nabla\cdot\left[ n\left( \nabla\phi_{1}+\nabla\phi_{2}\right) \right] = 0.
\end{eqnarray}
Thus, for a steady state, the phase gradients have to satisfy $\nabla\phi_{1}=-\nabla\phi_{2}$, which implies that the two pseudo-spin components acquire opposite velocities, and the total rotation of polariton BEC is zero. 
This solution is consistent with the expectation that a weak Raman coupling is insufficient to drive a rotation in polariton BEC.

\begin{figure*}
	\centering
	\includegraphics[width=1\textwidth]{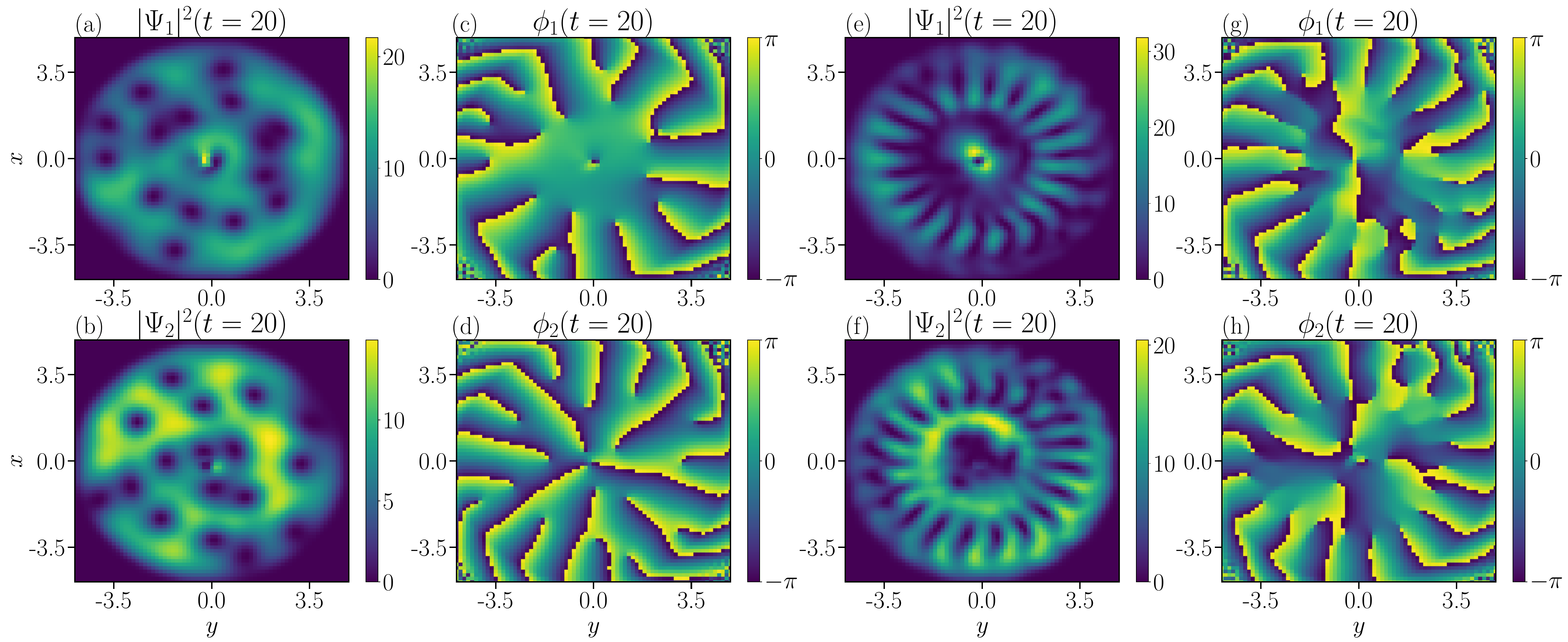}
	\caption{(Color online) 
		The density distributions $|\Psi_{1,2}|^2$ in (a)-(b) and (e)-(f), and corresponding phase distributions $\phi_{1,2}$ (c)-(d)  and (g)-(h) of two-component polariton BEC obtained by numerically solving Eqs.~(\ref{eqn:DGPE1})-(\ref{eqn:DGPE3}) with the different Raman coupling strength $\tilde{\Omega}^{\prime}_R= 1$ (first-second columns), $100$ (third-fourth columns) at $t = 20$. Other parameters are $\Omega_{v}=1$ and  $g^{\prime} = 0.5$.} 
	\label{fig2}
\end{figure*}    

When the Raman coupling strength $\tilde{\Omega}^{\prime}_R$ is increased, we must solve the coupled equations in Eq.~(\ref{eqn:phase_part11}) self-consistently. A stable solution can be easily found when $\phi_{1}-\phi_{2}=k\pi (k\in \mathbb{Z})$ is satisfied, and the velocities of the two components are $v_{1,2}=\nabla\phi_{1,2}=\pm\mathbf{l}\times\mathbf{r}/r^2 (r\neq 0)$. This solution implies that the polariton BEC can build a quantized circulation carrying a phase winding $\oint\nabla\phi_{1,2}\cdot \rm d$$\mathbf{s}=\pm2\pi l$ around the singularity in polariton cloud, which are similar to that in a polariton BEC pumped by a single LG beam~\cite{kwon2019}. When $\phi_{1}-\phi_{2} \neq k\pi$, a steady state can also exist if the polariton density vanishes $[n_{1,2}(r=0)=0]$ at the center of polariton cloud where the velocities diverge, which corresponds to the appearance of a vortex. The density far from the center $n_{1,2}(r\rightarrow\infty)$ can be finite because $\Omega_R(r\rightarrow\infty)=0$. Thus, a single quantized vortex with opposite phase winding can be stable as a steady state in a two-component polariton BEC with SOAMC and an infinite-size circular pumping beam, which is similar to the vortex-antivortex-pair phase discovered as a ground state of the Bose gas with SOAMC~\cite{Hui2020}.

When the circular pumping beam is of a finite radius with $P(r)=\eta P_{th}\Theta(R_p-r)$, the vortex instability induced by SOAMC will also be affected by the finite-size effect.
To this aim, we replace the pumping terms on the right-hand side of Eqs.~(\ref{eqn:DGPE1})-(\ref{eqn:DGPE2}) with an equivalently external rotating term $2\Omega_{v} L^{\prime}_{z}$ to investigate the steady-state solutions, where effective rotation frequency  $\Omega_{v}$ is related to the radius $R_p$ of the circular pump~\cite{Keeling2008}. With that, we obtain 
\begin{eqnarray}
	\label{eqn:phase_part21}
	\nabla&\cdot&\left[ n_{1}\left( \nabla\phi_{1}-\frac{\mathbf{l}}{r^2}\times\mathbf{r}-\mathbf{\Omega}_{v}\times\mathbf{r}\right) \right]\nonumber\\ 
	&=&-\Omega_{R}(r)\sqrt{n_{1}n_{2}}\sin (\phi_{1}-\phi_{2}), \nonumber\\
	\\	
	\nabla&\cdot&\left[ n_{2}\left( \nabla\phi_{2}+\frac{\mathbf{l}}{r^2}\times\mathbf{r}-\mathbf{\Omega}_{v}\times\mathbf{r}\right) \right]\nonumber\\ 
	&=&\Omega_{R}(r)\sqrt{n_{1}n_{2}}\sin (\phi_{1}-\phi_{2}),\nonumber	
\end{eqnarray}
where $\mathbf{\Omega}_{v}=\Omega_{v}\hat{z}$.

If the phase condition $\phi_{1}-\phi_{2}=k\pi$ $(k\in \mathbb{Z})$ is satisfied, a steady-state solution of vortex lattices can be obtained with  velocities $v_{1,2}\approx\mathbf{\Omega}_{v}\times\mathbf{r}\pm\mathbf{l}\times\mathbf{r}/r^2$ $(r\neq0)$. The number of vortices in each component is $\Omega_{v}R^{2}_{p}\pm l$. In addition, a quantized vortex still appears at the center with diverging velocities. When $\phi_{1}-\phi_{2}\neq k\pi$, vortices can exist at the trap center $r=0$ since the Raman coupling $\Omega_R(r=0)=0$. However, at other positions with non-zero $\Omega_R(r)$, stable vortex solutions can only be found with $n_1 = 0$ or $n_2 = 0$, i.e., the polariton BEC is fully polarized with the other component completely depleted. Furthermore, it is worth noting that the total OAM of polariton BEC is conserved. Thus, when a polariton BEC with SOAMC is subjected to a finite-size circular pumping beam, the center vortex-antivortex pair induced by SOAMC is still robust, while the vortex lattices steered by the boundary effect of pumping beam can be diminished.

\section{Dynamical evolution}
\label{sec:numerical}

\begin{figure*}
	\centering
	\includegraphics[width=1\textwidth]{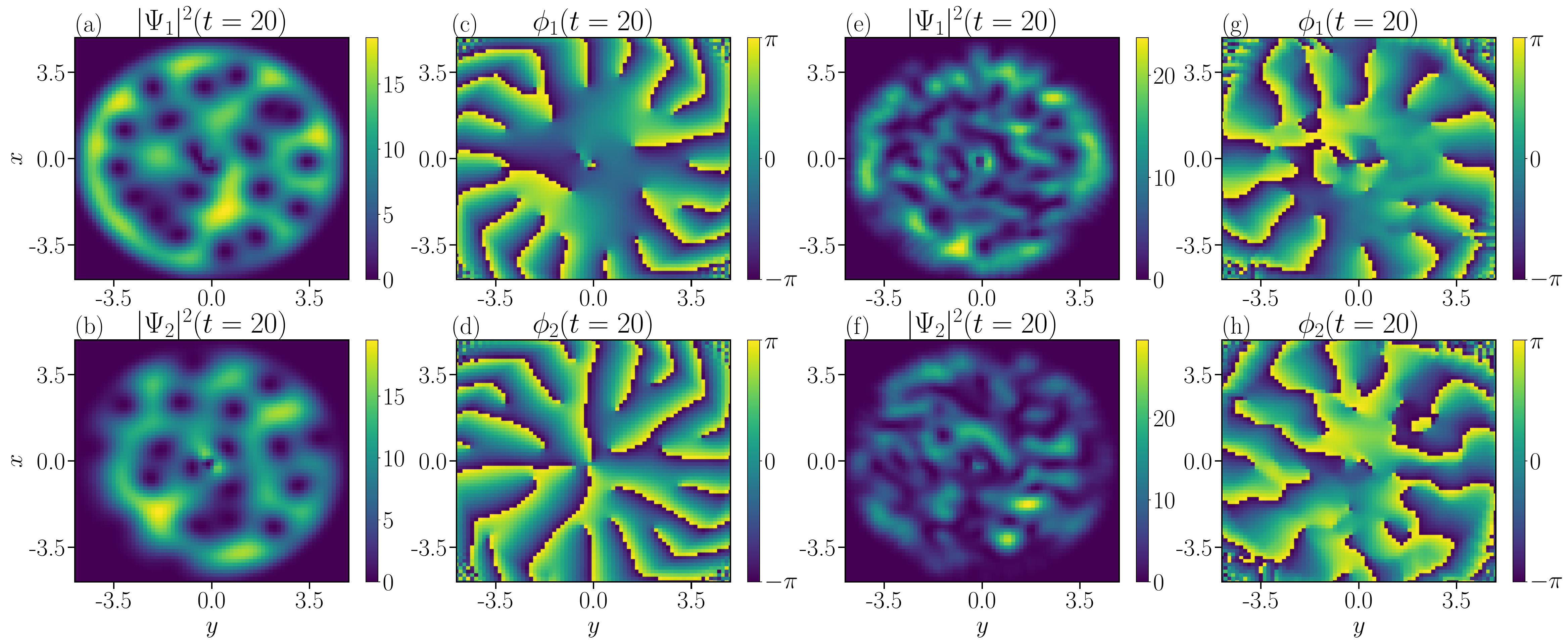}
	\caption{(Color online) 
		The density distributions $|\Psi_{1,2}|^2$ in (a)-(b) and (e)-(f), and corresponding phase distributions $\phi_{1,2}$ (c)-(d)  and (g)-(h) of two-component polariton BEC obtained by numerically solving Eqs.~(\ref{eqn:DGPE1})-(\ref{eqn:DGPE3}) with same parameter as Fig.~\ref{fig2}, except for $g^{\prime} = 0.1$}. 
	\label{fig3}
\end{figure*}    

To examine whether the steady-state solutions can actually be established in the dynamical evolution, in this section we investigate the stability and dynamical evolution of vortices by the time-dependent Eqs.~(\ref{eqn:DGPE1})-(\ref{eqn:DGPE3}). 
As presented in the Supplemental material,  we use the split-step Fourier method~\cite{bao2006efficient} for Eqs.~(\ref{eqn:DGPE1})-(\ref{eqn:DGPE2}) and the Runge-Kutta method for Eq.~(\ref{eqn:DGPE3}). Due to the presence of pumping and decay, the formation of a dynamical steady state usually takes long time for our present model within an inhomogeneous trapping potential. Thus, we adopt the conventional treatment of introducing an equivalently weak rotation term $-2\Omega_{v} L^{\prime}_{z}$ on the right-hand side of Eqs.~(\ref{eqn:DGPE1})-(\ref{eqn:DGPE2}) to reduce evolution time required for a dynamical steady state~\cite{Keeling2008}. The initial state is chosen in the form of a Thomas-Fermi distribution,
\begin{equation}
	\label{initial_TF}
	\Psi_{1,2}(t=0)=
	\left\{
	\begin{array}{ccc}
		\sqrt{\frac{3\alpha g^{\prime} - 2\beta r^2}{2\beta g^{\prime}}}&,&r<R_{TF},\\
		\\
		0&,&r\ge R_{TF},
	\end{array}
	\right.
\end{equation}
where $R_{TF} = \sqrt{3g^{\prime}\alpha/2\beta}$ is the Thomas-Fermi radius~\cite{Keeling2008} with the effective pumping rate $\alpha = P(r)R_{e}^{\prime}/2\gamma^{\prime}_e - \gamma^{\prime}_p/2$, and the effective rate of saturation loss $\beta=P(r)R_{e}^{\prime 2}/2\gamma^{\prime 2}_e$. In order to get closer to the steady-state solutions within a finite evolution time, we choose appropriate parameters to satisfy the adiabatic approximation conditions $\gamma^{\prime}_{e}\gg\gamma^{\prime}_{p}$ and $\gamma^{\prime}_{e}\gg P(r)R^{\prime}_{e}/2\gamma^{\prime}_{p}$. In the following numerical calculation, we use the dimensionless parameters $g^{\prime}_{12}=0.5g^{\prime}$, $g^{\prime}_e = 2g^{\prime}$, $\gamma^{\prime}_p=110$, $\gamma^{\prime}_e=100\gamma^{\prime}_p$, $R^{\prime}_{e}=0.51\gamma^{\prime}_p$, $\eta=1.08$, $\omega^{\prime} =7$, and $R_P=5$, which are of typical values in semiconductor microcavities~\cite{Georgios2011}.

In Fig.~\ref{fig1}, we present the results of evolution for a long enough time for the case of weak SOAMC with Raman coupling strength $\tilde{\Omega}^{\prime}_R=1$ and interaction $g^{\prime} = 1$. Vortex configurations at different moments of time are depicted in both the density and phase distributions with the initial condition $\Psi_{1,2}^0$ of Eq.~(\ref{initial_TF}). 
The location and number of vortices in the two components are clearly different and vary with time,
without showing any steady structure for the longest time we have achieved. 

In atomic BEC, the loss of particles may cause vortices to spiral towards regions where they vanish~\cite{Fetter2009,Rooney2010}. Spiraling vortices have also been observed in non-equilibrium polariton BECs, such as those pumped by a broad super-Gaussian beam with a flat-top~\cite{Sedov2021}, as well as for continuous-wave pump~\cite{Xuekai2020}. In the present case of a circular beam $P(r)=\eta P_{th}\Theta(R_p-r)$, spiraling vortices are discovered from the spiraling phase distributions. 
Moreover, two phase singularities with opposite winding located respectively at the center of two-component polariton clouds, as depicted by red rings in Figs.~\ref{fig1}(b)-\ref{fig1}(c) and directional rings in Figs.~\ref{fig1}(d)-\ref{fig1}(e), in consistence with the expectation of Eqs.~(\ref{eqn:phase_part21}). In summary, when the Raman coupling strength $\tilde{\Omega}^{\prime}_R=1$ is weak, the effective rotation $-2\Omega_{v} L^{\prime}_{z}$ dominates the dynamical evolution of vortices in polariton BEC, while the angular momentum exchange induced by SOAMC facilitates the stable lattice structure and destabilizes vortices.

For comparison, we also show the steady-state density profiles of two components in Fig.~\ref{fig1}(a), which are obtained by introducing a chemical potential $\mu$ and substituting $i\partial_t \Psi=\mu\Psi$ into Eqs.~(\ref{eqn:DGPE1})-(\ref{eqn:DGPE2}), where $\Psi = \left(\Psi_1,\Psi_2\right)$. Under the adiabatic approximation, we get 
\begin{eqnarray}
	\mu \Psi_{1}&=&\left[-\nabla^2_{r}+\tilde{V}_{1}(\mathbf{r},t)+\frac{(L^{\prime}_{z}-l)^2}{r^2}\right]\Psi_{1} \nonumber\\
	&&+\Omega_{R}(r)\Psi_{2}, \label{eqn:stead1}\\
	\
	\mu\Psi_{2}&=&\left[-\nabla^2_{r}+\tilde{V}_{2}(\mathbf{r},t)+\frac{(L^{\prime}_{z}+l)^2}{r^2}\right]\Psi_{2} \nonumber\\
	&&+\Omega_{R}(r)\Psi_{1}, \label{eqn:stead2}
\end{eqnarray}
where the effective potential $\tilde{V}_{j}(\mathbf{r},t) = r^2 + g^{\prime}  |\Psi_j|^2 + g^{\prime}_{12} |\Psi_{3-j}|^2 + i(\alpha-\beta (|\Psi_1|^2+|\Psi_2|^2))~(j = 1, 2)$. The chemical potential $\mu$ is determined by the balance of gain and loss~\cite{Keeling2008}. Based on the analysis of the previous section, we can safely assume that $\Psi$ has azimuthal symmetry and that $\phi_1 = \phi_2 = 0$. Then we use the Newton-Raphson method to calculate Eqs.~(\ref{eqn:stead1})-(\ref{eqn:stead2}) and scan various $\mu$ to obtain a steady-state solution. The results show that polariton BEC can be stabilized within the regime of $r<R_p$. 
When the SOAMC strength is enhanced, as shown the dashed line in Fig.~\ref{fig1}(a) with $\tilde{\Omega}^{\prime}_R=100$, steady-state solution of condensate can still be found, where polariton cloud has a smaller size compared to the case of $\tilde{\Omega}^{\prime}_R=1$.

In Fig.~\ref{fig2}, we present the vortex configuration in the dynamical evolution of polariton BEC with weak interaction $g^{\prime} = 0.5$. For the weak Raman coupling of SOAMC (first and second columns in Fig.~\ref{fig2}), comparing with vortices distribution in Figs.~\ref{fig1}(b)-\ref{fig1}(i) with $g^{\prime} = 1$, the boundary instability effect induced by the finite-size circular pump is more prominent. More vortices can spiral from the edge into polariton clouds with weak repulsive interaction. When the Raman coupling of SOAMC is increased,
we find that individual vortices can no longer be clearly recognized from density distributions [Figs.~\ref{fig2}(e)-(f)]. The existence of vortices can only be witnessed from the spiraling singularities in phase distributions of Fig.~\ref{fig2}(g)-(h). 
This observation can be understood by noticing that the spatially dependent Raman coupling $\Omega_R(r)$ destroys the translational symmetry and has the tendency to compromise any approximately uniform background density in a polariton cloud. In addition, strong Raman coupling shrinks the size of polariton distribution, which further destroys the stability of vortex lattices, allowing vortices to move freely into the condensate, except for the pair located at the center.

Finally, we discuss the effect of polariton interaction on vortex configuration. As shown in the Supplement material, in the absence of SOAMC the vortex configuration is regular and presents rotation symmetry in the case of weak repulsive interaction. A stronger interaction tends to disrupt the inherent symmetry of the vortex lattices. On the other hand, strong interaction helps to build a stable edge of the polariton cloud, preventing the emergence of vortices from the edge and resulting in fewer vortices. In the presence of SOAMC, however, the Raman coupling of SOAMC and boundary instability effect dominate the vortices distribution of polariton BEC, where the dynamical evolution of vortices is completely chaotic as vortices constantly spiral in the polariton clouds, as depicted in Fig.~\ref{fig3} with $g^{\prime} = 0.1$. Consequently, the vortices tend to disrupt the polariton BEC, which can be observed in both the density and phase distributions.

\section{CONCLUSION}
\label{sec:summary}
We propose to realize synthetic spin-orbital-angular-momentum coupling (SOAMC) in a two-component polariton BEC within a heterostructure of quantum well layers sandwiched by reflectors. By applying a pair of Laguerre-Gaussian beams to induce a Raman transition between different polariton branches, a pseudo-spin flip can be realized between the TE and TM modes of the lower polariton branch, accompanied by a change in orbital angular momentum. We derive the time-dependent open-dissipative Gross-Pitaevskii equation from single particle Hamiltonian, and investigate the stabilities of vortices and vortex lattices in the presence of a finite-size circular pump. 
We find that the presence of SOAMC can induce a pair of vortices located at the center of the polariton clouds, which remain stable as long as the polariton BEC exists. However, for the lattice configuration of other vortices, we conclude that while a repulsive interaction tends to stabilize vortex configuration in the competition between pump and decay of polariton BEC, the Raman coupling of SOAMC usually plays a dominant role by breaking the translational symmetry and causing disorder of vortices configuration. As a result, the vortex configurations are only stable with weak Raman coupling and strong interaction. When the Raman coupling strength increases and interaction decreases, the vortices spiraling in from the edge tend to disrupt polariton BEC.


\begin{acknowledgements}
This work is supported by the Natural Science Foundation of Zhejiang Province (Grant No. LQ22A040006), the National Natural Science Foundation of China (Grant Nos. 12074428, 92265208), and the National Key R$\&$D Program of China (Grant No. 2022YFA1405301).
\end{acknowledgements}


\begin{thebibliography}{64}%
\makeatletter
\providecommand \@ifxundefined [1]{%
 \@ifx{#1\undefined}
}%
\providecommand \@ifnum [1]{%
 \ifnum #1\expandafter \@firstoftwo
 \else \expandafter \@secondoftwo
 \fi
}%
\providecommand \@ifx [1]{%
 \ifx #1\expandafter \@firstoftwo
 \else \expandafter \@secondoftwo
 \fi
}%
\providecommand \natexlab [1]{#1}%
\providecommand \enquote  [1]{``#1''}%
\providecommand \bibnamefont  [1]{#1}%
\providecommand \bibfnamefont [1]{#1}%
\providecommand \citenamefont [1]{#1}%
\providecommand \href@noop [0]{\@secondoftwo}%
\providecommand \href [0]{\begingroup \@sanitize@url \@href}%
\providecommand \@href[1]{\@@startlink{#1}\@@href}%
\providecommand \@@href[1]{\endgroup#1\@@endlink}%
\providecommand \@sanitize@url [0]{\catcode `\\12\catcode `\$12\catcode
  `\&12\catcode `\#12\catcode `\^12\catcode `\_12\catcode `\%12\relax}%
\providecommand \@@startlink[1]{}%
\providecommand \@@endlink[0]{}%
\providecommand \url  [0]{\begingroup\@sanitize@url \@url }%
\providecommand \@url [1]{\endgroup\@href {#1}{\urlprefix }}%
\providecommand \urlprefix  [0]{URL }%
\providecommand \Eprint [0]{\href }%
\providecommand \doibase [0]{http://dx.doi.org/}%
\providecommand \selectlanguage [0]{\@gobble}%
\providecommand \bibinfo  [0]{\@secondoftwo}%
\providecommand \bibfield  [0]{\@secondoftwo}%
\providecommand \translation [1]{[#1]}%
\providecommand \BibitemOpen [0]{}%
\providecommand \bibitemStop [0]{}%
\providecommand \bibitemNoStop [0]{.\EOS\space}%
\providecommand \EOS [0]{\spacefactor3000\relax}%
\providecommand \BibitemShut  [1]{\csname bibitem#1\endcsname}%
\let\auto@bib@innerbib\@empty
\bibitem [{\citenamefont {Essmann}\ and\ \citenamefont
  {Tr\"{a}uble}(1967)}]{ESSMANN1967526}%
  \BibitemOpen
  \bibfield  {author} {\bibinfo {author} {\bibfnamefont {U.}~\bibnamefont
  {Essmann}}\ and\ \bibinfo {author} {\bibfnamefont {H.}~\bibnamefont
  {Tr\"{a}uble}},\ }\href {\doibase
  https://doi.org/10.1016/0375-9601(67)90819-5} {\bibfield  {journal} {\bibinfo
   {journal} {Phys. Lett. A}\ }\textbf {\bibinfo {volume} {24}},\ \bibinfo
  {pages} {526} (\bibinfo {year} {1967})}\BibitemShut {NoStop}%
\bibitem [{\citenamefont {Williams}\ and\ \citenamefont
  {Packard}(1974)}]{Williams1974}%
  \BibitemOpen
  \bibfield  {author} {\bibinfo {author} {\bibfnamefont {G.~A.}\ \bibnamefont
  {Williams}}\ and\ \bibinfo {author} {\bibfnamefont {R.~E.}\ \bibnamefont
  {Packard}},\ }\href {\doibase 10.1103/PhysRevLett.33.280} {\bibfield
  {journal} {\bibinfo  {journal} {Phys. Rev. Lett.}\ }\textbf {\bibinfo
  {volume} {33}},\ \bibinfo {pages} {280} (\bibinfo {year} {1974})}\BibitemShut
  {NoStop}%
\bibitem [{\citenamefont {Yarmchuk}\ \emph {et~al.}(1979)\citenamefont
  {Yarmchuk}, \citenamefont {Gordon},\ and\ \citenamefont
  {Packard}}]{Yarmchuk1979}%
  \BibitemOpen
  \bibfield  {author} {\bibinfo {author} {\bibfnamefont {E.~J.}\ \bibnamefont
  {Yarmchuk}}, \bibinfo {author} {\bibfnamefont {M.~J.~V.}\ \bibnamefont
  {Gordon}}, \ and\ \bibinfo {author} {\bibfnamefont {R.~E.}\ \bibnamefont
  {Packard}},\ }\href {\doibase 10.1103/PhysRevLett.43.214} {\bibfield
  {journal} {\bibinfo  {journal} {Phys. Rev. Lett.}\ }\textbf {\bibinfo
  {volume} {43}},\ \bibinfo {pages} {214} (\bibinfo {year} {1979})}\BibitemShut
  {NoStop}%
\bibitem [{\citenamefont {Madison}\ \emph {et~al.}(2000)\citenamefont
  {Madison}, \citenamefont {Chevy}, \citenamefont {Wohlleben},\ and\
  \citenamefont {Dalibard}}]{Madison2000}%
  \BibitemOpen
  \bibfield  {author} {\bibinfo {author} {\bibfnamefont {K.~W.}\ \bibnamefont
  {Madison}}, \bibinfo {author} {\bibfnamefont {F.}~\bibnamefont {Chevy}},
  \bibinfo {author} {\bibfnamefont {W.}~\bibnamefont {Wohlleben}}, \ and\
  \bibinfo {author} {\bibfnamefont {J.}~\bibnamefont {Dalibard}},\ }\href
  {\doibase 10.1103/PhysRevLett.84.806} {\bibfield  {journal} {\bibinfo
  {journal} {Phys. Rev. Lett.}\ }\textbf {\bibinfo {volume} {84}},\ \bibinfo
  {pages} {806} (\bibinfo {year} {2000})}\BibitemShut {NoStop}%
\bibitem [{\citenamefont {Abo-Shaeer}\ \emph {et~al.}(2001)\citenamefont
  {Abo-Shaeer}, \citenamefont {Raman}, \citenamefont {Vogels},\ and\
  \citenamefont {Ketterle}}]{abo2001observation}%
  \BibitemOpen
  \bibfield  {author} {\bibinfo {author} {\bibfnamefont {J.~R.}\ \bibnamefont
  {Abo-Shaeer}}, \bibinfo {author} {\bibfnamefont {C.}~\bibnamefont {Raman}},
  \bibinfo {author} {\bibfnamefont {J.~M.}\ \bibnamefont {Vogels}}, \ and\
  \bibinfo {author} {\bibfnamefont {W.}~\bibnamefont {Ketterle}},\ }\href
  {\doibase 10.1126/science.1060182} {\bibfield  {journal} {\bibinfo  {journal}
  {Science}\ }\textbf {\bibinfo {volume} {292}},\ \bibinfo {pages} {476}
  (\bibinfo {year} {2001})}\BibitemShut {NoStop}%
\bibitem [{\citenamefont {Zhen}\ \emph {et~al.}(2014)\citenamefont {Zhen},
  \citenamefont {Hsu}, \citenamefont {Lu}, \citenamefont {Stone},\ and\
  \citenamefont {Solja\ifmmode \check{c}\else
  \v{c}\fi{}i\ifmmode~\acute{c}\else \'{c}\fi{}}}]{Zhen2014}%
  \BibitemOpen
  \bibfield  {author} {\bibinfo {author} {\bibfnamefont {B.}~\bibnamefont
  {Zhen}}, \bibinfo {author} {\bibfnamefont {C.~W.}\ \bibnamefont {Hsu}},
  \bibinfo {author} {\bibfnamefont {L.}~\bibnamefont {Lu}}, \bibinfo {author}
  {\bibfnamefont {A.~D.}\ \bibnamefont {Stone}}, \ and\ \bibinfo {author}
  {\bibfnamefont {M.}~\bibnamefont {Solja\ifmmode \check{c}\else
  \v{c}\fi{}i\ifmmode~\acute{c}\else \'{c}\fi{}}},\ }\href {\doibase
  10.1103/PhysRevLett.113.257401} {\bibfield  {journal} {\bibinfo  {journal}
  {Phys. Rev. Lett.}\ }\textbf {\bibinfo {volume} {113}},\ \bibinfo {pages}
  {257401} (\bibinfo {year} {2014})}\BibitemShut {NoStop}%
\bibitem [{\citenamefont {Zhang}\ \emph {et~al.}(2018)\citenamefont {Zhang},
  \citenamefont {Chen}, \citenamefont {Liu}, \citenamefont {Hsu}, \citenamefont
  {Wang}, \citenamefont {Guan}, \citenamefont {Liu}, \citenamefont {Shi},
  \citenamefont {Lu},\ and\ \citenamefont {Zi}}]{Zijian2018}%
  \BibitemOpen
  \bibfield  {author} {\bibinfo {author} {\bibfnamefont {Y.}~\bibnamefont
  {Zhang}}, \bibinfo {author} {\bibfnamefont {A.}~\bibnamefont {Chen}},
  \bibinfo {author} {\bibfnamefont {W.}~\bibnamefont {Liu}}, \bibinfo {author}
  {\bibfnamefont {C.~W.}\ \bibnamefont {Hsu}}, \bibinfo {author} {\bibfnamefont
  {B.}~\bibnamefont {Wang}}, \bibinfo {author} {\bibfnamefont {F.}~\bibnamefont
  {Guan}}, \bibinfo {author} {\bibfnamefont {X.}~\bibnamefont {Liu}}, \bibinfo
  {author} {\bibfnamefont {L.}~\bibnamefont {Shi}}, \bibinfo {author}
  {\bibfnamefont {L.}~\bibnamefont {Lu}}, \ and\ \bibinfo {author}
  {\bibfnamefont {J.}~\bibnamefont {Zi}},\ }\href {\doibase
  10.1103/PhysRevLett.120.186103} {\bibfield  {journal} {\bibinfo  {journal}
  {Phys. Rev. Lett.}\ }\textbf {\bibinfo {volume} {120}},\ \bibinfo {pages}
  {186103} (\bibinfo {year} {2018})}\BibitemShut {NoStop}%
\bibitem [{\citenamefont {Lagoudakis}\ \emph {et~al.}(2008)\citenamefont
  {Lagoudakis}, \citenamefont {Wouters}, \citenamefont {Richard}, \citenamefont
  {Baas}, \citenamefont {Carusotto}, \citenamefont {Andr{\'e}}, \citenamefont
  {Dang},\ and\ \citenamefont {Deveaud-Pl{\'e}dran}}]{lagoudakis2008}%
  \BibitemOpen
  \bibfield  {author} {\bibinfo {author} {\bibfnamefont {K.~G.}\ \bibnamefont
  {Lagoudakis}}, \bibinfo {author} {\bibfnamefont {M.}~\bibnamefont {Wouters}},
  \bibinfo {author} {\bibfnamefont {M.}~\bibnamefont {Richard}}, \bibinfo
  {author} {\bibfnamefont {A.}~\bibnamefont {Baas}}, \bibinfo {author}
  {\bibfnamefont {I.}~\bibnamefont {Carusotto}}, \bibinfo {author}
  {\bibfnamefont {R.}~\bibnamefont {Andr{\'e}}}, \bibinfo {author}
  {\bibfnamefont {L.~S.}\ \bibnamefont {Dang}}, \ and\ \bibinfo {author}
  {\bibfnamefont {B.}~\bibnamefont {Deveaud-Pl{\'e}dran}},\ }\href
  {https://doi.org/10.1038/nphys1051} {\bibfield  {journal} {\bibinfo
  {journal} {Nat. Phys.}\ }\textbf {\bibinfo {volume} {4}},\ \bibinfo {pages}
  {706} (\bibinfo {year} {2008})}\BibitemShut {NoStop}%
\bibitem [{\citenamefont {Lagoudakis}\ \emph {et~al.}(2009)\citenamefont
  {Lagoudakis}, \citenamefont {Ostatnick\'{y}}, \citenamefont {Kavokin},
  \citenamefont {Rubo}, \citenamefont {Andr\'{e}},\ and\ \citenamefont
  {Deveaud-Pl\'{e}dran}}]{Lagoudakis2009}%
  \BibitemOpen
  \bibfield  {author} {\bibinfo {author} {\bibfnamefont {K.~G.}\ \bibnamefont
  {Lagoudakis}}, \bibinfo {author} {\bibfnamefont {T.}~\bibnamefont
  {Ostatnick\'{y}}}, \bibinfo {author} {\bibfnamefont {A.~V.}\ \bibnamefont
  {Kavokin}}, \bibinfo {author} {\bibfnamefont {Y.~G.}\ \bibnamefont {Rubo}},
  \bibinfo {author} {\bibfnamefont {R.}~\bibnamefont {Andr\'{e}}}, \ and\
  \bibinfo {author} {\bibfnamefont {B.}~\bibnamefont {Deveaud-Pl\'{e}dran}},\
  }\href {\doibase 10.1126/science.1177980} {\bibfield  {journal} {\bibinfo
  {journal} {Science}\ }\textbf {\bibinfo {volume} {326}},\ \bibinfo {pages}
  {974} (\bibinfo {year} {2009})}\BibitemShut {NoStop}%
\bibitem [{\citenamefont {Su}\ \emph {et~al.}(2021)\citenamefont {Su},
  \citenamefont {Ghosh}, \citenamefont {Liew},\ and\ \citenamefont
  {Xiong}}]{QihuaXiong2021}%
  \BibitemOpen
  \bibfield  {author} {\bibinfo {author} {\bibfnamefont {R.}~\bibnamefont
  {Su}}, \bibinfo {author} {\bibfnamefont {S.}~\bibnamefont {Ghosh}}, \bibinfo
  {author} {\bibfnamefont {T.~C.~H.}\ \bibnamefont {Liew}}, \ and\ \bibinfo
  {author} {\bibfnamefont {Q.}~\bibnamefont {Xiong}},\ }\href {\doibase
  10.1126/sciadv.abf8049} {\bibfield  {journal} {\bibinfo  {journal} {Sci.
  Adv.}\ }\textbf {\bibinfo {volume} {7}},\ \bibinfo {pages} {eabf8049}
  (\bibinfo {year} {2021})}\BibitemShut {NoStop}%
\bibitem [{\citenamefont {Real}\ \emph {et~al.}(2021)\citenamefont {Real},
  \citenamefont {Carlon~Zambon}, \citenamefont {St-Jean}, \citenamefont
  {Sagnes}, \citenamefont {Lema\^{\i}tre}, \citenamefont {Le~Gratiet},
  \citenamefont {Harouri}, \citenamefont {Ravets}, \citenamefont {Bloch},\ and\
  \citenamefont {Amo}}]{Real2021}%
  \BibitemOpen
  \bibfield  {author} {\bibinfo {author} {\bibfnamefont {B.}~\bibnamefont
  {Real}}, \bibinfo {author} {\bibfnamefont {N.}~\bibnamefont {Carlon~Zambon}},
  \bibinfo {author} {\bibfnamefont {P.}~\bibnamefont {St-Jean}}, \bibinfo
  {author} {\bibfnamefont {I.}~\bibnamefont {Sagnes}}, \bibinfo {author}
  {\bibfnamefont {A.}~\bibnamefont {Lema\^{\i}tre}}, \bibinfo {author}
  {\bibfnamefont {L.}~\bibnamefont {Le~Gratiet}}, \bibinfo {author}
  {\bibfnamefont {A.}~\bibnamefont {Harouri}}, \bibinfo {author} {\bibfnamefont
  {S.}~\bibnamefont {Ravets}}, \bibinfo {author} {\bibfnamefont
  {J.}~\bibnamefont {Bloch}}, \ and\ \bibinfo {author} {\bibfnamefont
  {A.}~\bibnamefont {Amo}},\ }\href {\doibase 10.1103/PhysRevResearch.3.043161}
  {\bibfield  {journal} {\bibinfo  {journal} {Phys. Rev. Research}\ }\textbf
  {\bibinfo {volume} {3}},\ \bibinfo {pages} {043161} (\bibinfo {year}
  {2021})}\BibitemShut {NoStop}%
\bibitem [{\citenamefont {Deng}\ \emph {et~al.}(2010)\citenamefont {Deng},
  \citenamefont {Haug},\ and\ \citenamefont {Yamamoto}}]{Deng2010}%
  \BibitemOpen
  \bibfield  {author} {\bibinfo {author} {\bibfnamefont {H.}~\bibnamefont
  {Deng}}, \bibinfo {author} {\bibfnamefont {H.}~\bibnamefont {Haug}}, \ and\
  \bibinfo {author} {\bibfnamefont {Y.}~\bibnamefont {Yamamoto}},\ }\href
  {\doibase 10.1103/RevModPhys.82.1489} {\bibfield  {journal} {\bibinfo
  {journal} {Rev. Mod. Phys.}\ }\textbf {\bibinfo {volume} {82}},\ \bibinfo
  {pages} {1489} (\bibinfo {year} {2010})}\BibitemShut {NoStop}%
\bibitem [{\citenamefont {Levinsen}\ \emph {et~al.}(2019)\citenamefont
  {Levinsen}, \citenamefont {Li},\ and\ \citenamefont {Parish}}]{Levinsen2019}%
  \BibitemOpen
  \bibfield  {author} {\bibinfo {author} {\bibfnamefont {J.}~\bibnamefont
  {Levinsen}}, \bibinfo {author} {\bibfnamefont {G.}~\bibnamefont {Li}}, \ and\
  \bibinfo {author} {\bibfnamefont {M.~M.}\ \bibnamefont {Parish}},\ }\href
  {\doibase 10.1103/PhysRevResearch.1.033120} {\bibfield  {journal} {\bibinfo
  {journal} {Phys. Rev. Research}\ }\textbf {\bibinfo {volume} {1}},\ \bibinfo
  {pages} {033120} (\bibinfo {year} {2019})}\BibitemShut {NoStop}%
\bibitem [{\citenamefont {Panico}\ \emph {et~al.}(2021)\citenamefont {Panico},
  \citenamefont {Macorini}, \citenamefont {Dominici}, \citenamefont
  {Gianfrate}, \citenamefont {Fieramosca}, \citenamefont {De~Giorgi},
  \citenamefont {Gigli}, \citenamefont {Sanvitto}, \citenamefont {Lanotte},\
  and\ \citenamefont {Ballarini}}]{Panico2021}%
  \BibitemOpen
  \bibfield  {author} {\bibinfo {author} {\bibfnamefont {R.}~\bibnamefont
  {Panico}}, \bibinfo {author} {\bibfnamefont {G.}~\bibnamefont {Macorini}},
  \bibinfo {author} {\bibfnamefont {L.}~\bibnamefont {Dominici}}, \bibinfo
  {author} {\bibfnamefont {A.}~\bibnamefont {Gianfrate}}, \bibinfo {author}
  {\bibfnamefont {A.}~\bibnamefont {Fieramosca}}, \bibinfo {author}
  {\bibfnamefont {M.}~\bibnamefont {De~Giorgi}}, \bibinfo {author}
  {\bibfnamefont {G.}~\bibnamefont {Gigli}}, \bibinfo {author} {\bibfnamefont
  {D.}~\bibnamefont {Sanvitto}}, \bibinfo {author} {\bibfnamefont {A.~S.}\
  \bibnamefont {Lanotte}}, \ and\ \bibinfo {author} {\bibfnamefont
  {D.}~\bibnamefont {Ballarini}},\ }\href {\doibase
  10.1103/PhysRevLett.127.047401} {\bibfield  {journal} {\bibinfo  {journal}
  {Phys. Rev. Lett.}\ }\textbf {\bibinfo {volume} {127}},\ \bibinfo {pages}
  {047401} (\bibinfo {year} {2021})}\BibitemShut {NoStop}%
\bibitem [{\citenamefont {Ardizzone}\ \emph {et~al.}(2022)\citenamefont
  {Ardizzone}, \citenamefont {Riminucci}, \citenamefont {Zanotti},
  \citenamefont {Gianfrate}, \citenamefont {Efthymiou-Tsironi}, \citenamefont
  {Su{\`a}rez-Forero}, \citenamefont {Todisco}, \citenamefont {De~Giorgi},
  \citenamefont {Trypogeorgos}, \citenamefont {Gigli}, \citenamefont {Baldwin},
  \citenamefont {Pfeiffer}, \citenamefont {Ballarini}, \citenamefont {Nguyen},
  \citenamefont {Gerace},\ and\ \citenamefont {Sanvitto}}]{Ardizzone2022}%
  \BibitemOpen
  \bibfield  {author} {\bibinfo {author} {\bibfnamefont {V.}~\bibnamefont
  {Ardizzone}}, \bibinfo {author} {\bibfnamefont {F.}~\bibnamefont
  {Riminucci}}, \bibinfo {author} {\bibfnamefont {S.}~\bibnamefont {Zanotti}},
  \bibinfo {author} {\bibfnamefont {A.}~\bibnamefont {Gianfrate}}, \bibinfo
  {author} {\bibfnamefont {M.}~\bibnamefont {Efthymiou-Tsironi}}, \bibinfo
  {author} {\bibfnamefont {D.~G.}\ \bibnamefont {Su{\`a}rez-Forero}}, \bibinfo
  {author} {\bibfnamefont {F.}~\bibnamefont {Todisco}}, \bibinfo {author}
  {\bibfnamefont {M.}~\bibnamefont {De~Giorgi}}, \bibinfo {author}
  {\bibfnamefont {D.}~\bibnamefont {Trypogeorgos}}, \bibinfo {author}
  {\bibfnamefont {G.}~\bibnamefont {Gigli}}, \bibinfo {author} {\bibfnamefont
  {K.}~\bibnamefont {Baldwin}}, \bibinfo {author} {\bibfnamefont
  {L.}~\bibnamefont {Pfeiffer}}, \bibinfo {author} {\bibfnamefont
  {D.}~\bibnamefont {Ballarini}}, \bibinfo {author} {\bibfnamefont {H.~S.}\
  \bibnamefont {Nguyen}}, \bibinfo {author} {\bibfnamefont {D.}~\bibnamefont
  {Gerace}}, \ and\ \bibinfo {author} {\bibfnamefont {D.}~\bibnamefont
  {Sanvitto}},\ }\href {https://doi.org/10.1038/s41586-022-04583-7} {\bibfield
  {journal} {\bibinfo  {journal} {Nature}\ }\textbf {\bibinfo {volume} {605}},\
  \bibinfo {pages} {447} (\bibinfo {year} {2022})}\BibitemShut {NoStop}%
\bibitem [{\citenamefont {Boulier}\ \emph {et~al.}(2015)\citenamefont
  {Boulier}, \citenamefont {Ter{\c{c}}as}, \citenamefont {Solnyshkov},
  \citenamefont {Glorieux}, \citenamefont {Giacobino}, \citenamefont
  {Malpuech},\ and\ \citenamefont {Bramati}}]{boulier2015}%
  \BibitemOpen
  \bibfield  {author} {\bibinfo {author} {\bibfnamefont {T.}~\bibnamefont
  {Boulier}}, \bibinfo {author} {\bibfnamefont {H.}~\bibnamefont
  {Ter{\c{c}}as}}, \bibinfo {author} {\bibfnamefont {D.}~\bibnamefont
  {Solnyshkov}}, \bibinfo {author} {\bibfnamefont {Q.}~\bibnamefont
  {Glorieux}}, \bibinfo {author} {\bibfnamefont {E.}~\bibnamefont {Giacobino}},
  \bibinfo {author} {\bibfnamefont {G.}~\bibnamefont {Malpuech}}, \ and\
  \bibinfo {author} {\bibfnamefont {A.}~\bibnamefont {Bramati}},\ }\href
  {https://doi.org/10.1038/srep09230} {\bibfield  {journal} {\bibinfo
  {journal} {Sci. Rep.}\ }\textbf {\bibinfo {volume} {5}},\ \bibinfo {pages}
  {1} (\bibinfo {year} {2015})}\BibitemShut {NoStop}%
\bibitem [{\citenamefont {Kwon}\ \emph {et~al.}(2019)\citenamefont {Kwon},
  \citenamefont {Oh}, \citenamefont {Gong}, \citenamefont {Kim}, \citenamefont
  {Kang}, \citenamefont {Kang}, \citenamefont {Song}, \citenamefont {Choi},\
  and\ \citenamefont {Cho}}]{kwon2019}%
  \BibitemOpen
  \bibfield  {author} {\bibinfo {author} {\bibfnamefont {M.-S.}\ \bibnamefont
  {Kwon}}, \bibinfo {author} {\bibfnamefont {B.~Y.}\ \bibnamefont {Oh}},
  \bibinfo {author} {\bibfnamefont {S.-H.}\ \bibnamefont {Gong}}, \bibinfo
  {author} {\bibfnamefont {J.-H.}\ \bibnamefont {Kim}}, \bibinfo {author}
  {\bibfnamefont {H.~K.}\ \bibnamefont {Kang}}, \bibinfo {author}
  {\bibfnamefont {S.}~\bibnamefont {Kang}}, \bibinfo {author} {\bibfnamefont
  {J.~D.}\ \bibnamefont {Song}}, \bibinfo {author} {\bibfnamefont
  {H.}~\bibnamefont {Choi}}, \ and\ \bibinfo {author} {\bibfnamefont {Y.-H.}\
  \bibnamefont {Cho}},\ }\href {\doibase 10.1103/PhysRevLett.122.045302}
  {\bibfield  {journal} {\bibinfo  {journal} {Phys. Rev. Lett.}\ }\textbf
  {\bibinfo {volume} {122}},\ \bibinfo {pages} {045302} (\bibinfo {year}
  {2019})}\BibitemShut {NoStop}%
\bibitem [{\citenamefont {Choi}\ \emph {et~al.}(2022)\citenamefont {Choi},
  \citenamefont {Park}, \citenamefont {Oh}, \citenamefont {Kwon}, \citenamefont
  {Park}, \citenamefont {Kang}, \citenamefont {Song}, \citenamefont {Ko},
  \citenamefont {Sun}, \citenamefont {Savenko}, \citenamefont {Cho},\ and\
  \citenamefont {Choi}}]{Choi2022}%
  \BibitemOpen
  \bibfield  {author} {\bibinfo {author} {\bibfnamefont {D.}~\bibnamefont
  {Choi}}, \bibinfo {author} {\bibfnamefont {M.}~\bibnamefont {Park}}, \bibinfo
  {author} {\bibfnamefont {B.~Y.}\ \bibnamefont {Oh}}, \bibinfo {author}
  {\bibfnamefont {M.-S.}\ \bibnamefont {Kwon}}, \bibinfo {author}
  {\bibfnamefont {S.~I.}\ \bibnamefont {Park}}, \bibinfo {author}
  {\bibfnamefont {S.}~\bibnamefont {Kang}}, \bibinfo {author} {\bibfnamefont
  {J.~D.}\ \bibnamefont {Song}}, \bibinfo {author} {\bibfnamefont
  {D.}~\bibnamefont {Ko}}, \bibinfo {author} {\bibfnamefont {M.}~\bibnamefont
  {Sun}}, \bibinfo {author} {\bibfnamefont {I.~G.}\ \bibnamefont {Savenko}},
  \bibinfo {author} {\bibfnamefont {Y.-H.}\ \bibnamefont {Cho}}, \ and\
  \bibinfo {author} {\bibfnamefont {H.}~\bibnamefont {Choi}},\ }\href {\doibase
  10.1103/PhysRevB.105.L060502} {\bibfield  {journal} {\bibinfo  {journal}
  {Phys. Rev. B}\ }\textbf {\bibinfo {volume} {105}},\ \bibinfo {pages}
  {L060502} (\bibinfo {year} {2022})}\BibitemShut {NoStop}%
\bibitem [{\citenamefont {Sanvitto}\ \emph {et~al.}(2010)\citenamefont
  {Sanvitto}, \citenamefont {Marchetti}, \citenamefont {Szyma{\'n}ska},
  \citenamefont {Tosi}, \citenamefont {Baudisch}, \citenamefont {Laussy},
  \citenamefont {Krizhanovskii}, \citenamefont {Skolnick}, \citenamefont
  {Marrucci}, \citenamefont {Lema\^{i}tre}, \citenamefont {Bloch},
  \citenamefont {Tejedor},\ and\ \citenamefont {Vi\~{n}a}}]{sanvitto2010}%
  \BibitemOpen
  \bibfield  {author} {\bibinfo {author} {\bibfnamefont {D.}~\bibnamefont
  {Sanvitto}}, \bibinfo {author} {\bibfnamefont {F.}~\bibnamefont {Marchetti}},
  \bibinfo {author} {\bibfnamefont {M.}~\bibnamefont {Szyma{\'n}ska}}, \bibinfo
  {author} {\bibfnamefont {G.}~\bibnamefont {Tosi}}, \bibinfo {author}
  {\bibfnamefont {M.}~\bibnamefont {Baudisch}}, \bibinfo {author}
  {\bibfnamefont {F.~P.}\ \bibnamefont {Laussy}}, \bibinfo {author}
  {\bibfnamefont {D.}~\bibnamefont {Krizhanovskii}}, \bibinfo {author}
  {\bibfnamefont {M.}~\bibnamefont {Skolnick}}, \bibinfo {author}
  {\bibfnamefont {L.}~\bibnamefont {Marrucci}}, \bibinfo {author}
  {\bibfnamefont {A.}~\bibnamefont {Lema\^{i}tre}}, \bibinfo {author}
  {\bibfnamefont {J.}~\bibnamefont {Bloch}}, \bibinfo {author} {\bibfnamefont
  {C.}~\bibnamefont {Tejedor}}, \ and\ \bibinfo {author} {\bibfnamefont
  {L.}~\bibnamefont {Vi\~{n}a}},\ }\href {\doibase 10.1038/nphys1668}
  {\bibfield  {journal} {\bibinfo  {journal} {Nat. Phys.}\ }\textbf {\bibinfo
  {volume} {6}},\ \bibinfo {pages} {527} (\bibinfo {year} {2010})}\BibitemShut
  {NoStop}%
\bibitem [{\citenamefont {Boulier}\ \emph {et~al.}(2016)\citenamefont
  {Boulier}, \citenamefont {Cancellieri}, \citenamefont {Sangouard},
  \citenamefont {Glorieux}, \citenamefont {Kavokin}, \citenamefont {Whittaker},
  \citenamefont {Giacobino},\ and\ \citenamefont {Bramati}}]{Boulier2016}%
  \BibitemOpen
  \bibfield  {author} {\bibinfo {author} {\bibfnamefont {T.}~\bibnamefont
  {Boulier}}, \bibinfo {author} {\bibfnamefont {E.}~\bibnamefont
  {Cancellieri}}, \bibinfo {author} {\bibfnamefont {N.~D.}\ \bibnamefont
  {Sangouard}}, \bibinfo {author} {\bibfnamefont {Q.}~\bibnamefont {Glorieux}},
  \bibinfo {author} {\bibfnamefont {A.~V.}\ \bibnamefont {Kavokin}}, \bibinfo
  {author} {\bibfnamefont {D.~M.}\ \bibnamefont {Whittaker}}, \bibinfo {author}
  {\bibfnamefont {E.}~\bibnamefont {Giacobino}}, \ and\ \bibinfo {author}
  {\bibfnamefont {A.}~\bibnamefont {Bramati}},\ }\href {\doibase
  10.1103/PhysRevLett.116.116402} {\bibfield  {journal} {\bibinfo  {journal}
  {Phys. Rev. Lett.}\ }\textbf {\bibinfo {volume} {116}},\ \bibinfo {pages}
  {116402} (\bibinfo {year} {2016})}\BibitemShut {NoStop}%
\bibitem [{\citenamefont {Hu}\ \emph {et~al.}(2020)\citenamefont {Hu},
  \citenamefont {Kim}, \citenamefont {Schneider}, \citenamefont {H\"ofling},\
  and\ \citenamefont {Deng}}]{Hu2020}%
  \BibitemOpen
  \bibfield  {author} {\bibinfo {author} {\bibfnamefont {J.}~\bibnamefont
  {Hu}}, \bibinfo {author} {\bibfnamefont {S.}~\bibnamefont {Kim}}, \bibinfo
  {author} {\bibfnamefont {C.}~\bibnamefont {Schneider}}, \bibinfo {author}
  {\bibfnamefont {S.}~\bibnamefont {H\"ofling}}, \ and\ \bibinfo {author}
  {\bibfnamefont {H.}~\bibnamefont {Deng}},\ }\href {\doibase
  10.1103/PhysRevApplied.14.044001} {\bibfield  {journal} {\bibinfo  {journal}
  {Phys. Rev. Applied}\ }\textbf {\bibinfo {volume} {14}},\ \bibinfo {pages}
  {044001} (\bibinfo {year} {2020})}\BibitemShut {NoStop}%
\bibitem [{\citenamefont {Panzarini}\ \emph {et~al.}(1999)\citenamefont
  {Panzarini}, \citenamefont {Andreani}, \citenamefont {Armitage},
  \citenamefont {Baxter}, \citenamefont {Skolnick}, \citenamefont {Astratov},
  \citenamefont {Roberts}, \citenamefont {Kavokin}, \citenamefont
  {Vladimirova},\ and\ \citenamefont {Kaliteevski}}]{Panzarini1999}%
  \BibitemOpen
  \bibfield  {author} {\bibinfo {author} {\bibfnamefont {G.}~\bibnamefont
  {Panzarini}}, \bibinfo {author} {\bibfnamefont {L.~C.}\ \bibnamefont
  {Andreani}}, \bibinfo {author} {\bibfnamefont {A.}~\bibnamefont {Armitage}},
  \bibinfo {author} {\bibfnamefont {D.}~\bibnamefont {Baxter}}, \bibinfo
  {author} {\bibfnamefont {M.~S.}\ \bibnamefont {Skolnick}}, \bibinfo {author}
  {\bibfnamefont {V.~N.}\ \bibnamefont {Astratov}}, \bibinfo {author}
  {\bibfnamefont {J.~S.}\ \bibnamefont {Roberts}}, \bibinfo {author}
  {\bibfnamefont {A.~V.}\ \bibnamefont {Kavokin}}, \bibinfo {author}
  {\bibfnamefont {M.~R.}\ \bibnamefont {Vladimirova}}, \ and\ \bibinfo {author}
  {\bibfnamefont {M.~A.}\ \bibnamefont {Kaliteevski}},\ }\href {\doibase
  10.1103/PhysRevB.59.5082} {\bibfield  {journal} {\bibinfo  {journal} {Phys.
  Rev. B}\ }\textbf {\bibinfo {volume} {59}},\ \bibinfo {pages} {5082}
  (\bibinfo {year} {1999})}\BibitemShut {NoStop}%
\bibitem [{\citenamefont {Kavokin}\ \emph {et~al.}(2005)\citenamefont
  {Kavokin}, \citenamefont {Malpuech},\ and\ \citenamefont
  {Glazov}}]{Kavokin2005}%
  \BibitemOpen
  \bibfield  {author} {\bibinfo {author} {\bibfnamefont {A.}~\bibnamefont
  {Kavokin}}, \bibinfo {author} {\bibfnamefont {G.}~\bibnamefont {Malpuech}}, \
  and\ \bibinfo {author} {\bibfnamefont {M.}~\bibnamefont {Glazov}},\ }\href
  {\doibase 10.1103/PhysRevLett.95.136601} {\bibfield  {journal} {\bibinfo
  {journal} {Phys. Rev. Lett.}\ }\textbf {\bibinfo {volume} {95}},\ \bibinfo
  {pages} {136601} (\bibinfo {year} {2005})}\BibitemShut {NoStop}%
\bibitem [{\citenamefont {Tassone}\ \emph {et~al.}(1992)\citenamefont
  {Tassone}, \citenamefont {Bassani},\ and\ \citenamefont
  {Andreani}}]{Tassone1992}%
  \BibitemOpen
  \bibfield  {author} {\bibinfo {author} {\bibfnamefont {F.}~\bibnamefont
  {Tassone}}, \bibinfo {author} {\bibfnamefont {F.}~\bibnamefont {Bassani}}, \
  and\ \bibinfo {author} {\bibfnamefont {L.~C.}\ \bibnamefont {Andreani}},\
  }\href {\doibase 10.1103/PhysRevB.45.6023} {\bibfield  {journal} {\bibinfo
  {journal} {Phys. Rev. B}\ }\textbf {\bibinfo {volume} {45}},\ \bibinfo
  {pages} {6023} (\bibinfo {year} {1992})}\BibitemShut {NoStop}%
\bibitem [{\citenamefont {Leyder}\ \emph {et~al.}(2007)\citenamefont {Leyder},
  \citenamefont {Romanelli}, \citenamefont {Karr}, \citenamefont {Giacobino},
  \citenamefont {Liew}, \citenamefont {Glazov}, \citenamefont {Kavokin},
  \citenamefont {Malpuech},\ and\ \citenamefont {Rramati}}]{Leyder2007}%
  \BibitemOpen
  \bibfield  {author} {\bibinfo {author} {\bibfnamefont {C.}~\bibnamefont
  {Leyder}}, \bibinfo {author} {\bibfnamefont {M.}~\bibnamefont {Romanelli}},
  \bibinfo {author} {\bibfnamefont {J.~P.}\ \bibnamefont {Karr}}, \bibinfo
  {author} {\bibfnamefont {E.}~\bibnamefont {Giacobino}}, \bibinfo {author}
  {\bibfnamefont {T.}~\bibnamefont {Liew}}, \bibinfo {author} {\bibfnamefont
  {M.~M.}\ \bibnamefont {Glazov}}, \bibinfo {author} {\bibfnamefont
  {A.}~\bibnamefont {Kavokin}}, \bibinfo {author} {\bibfnamefont
  {G.}~\bibnamefont {Malpuech}}, \ and\ \bibinfo {author} {\bibfnamefont
  {A.}~\bibnamefont {Rramati}},\ }\href {\doibase 10.1038/nphys676} {\bibfield
  {journal} {\bibinfo  {journal} {Nat. Phys.}\ }\textbf {\bibinfo {volume}
  {3}},\ \bibinfo {pages} {628} (\bibinfo {year} {2007})}\BibitemShut {NoStop}%
\bibitem [{\citenamefont {Kammann}\ \emph {et~al.}(2012)\citenamefont
  {Kammann}, \citenamefont {Liew}, \citenamefont {Ohadi}, \citenamefont
  {Cilibrizzi}, \citenamefont {Tsotsis}, \citenamefont {Hatzopoulos},
  \citenamefont {Savvidis}, \citenamefont {Kavokin},\ and\ \citenamefont
  {Lagoudakis}}]{Kammann2012}%
  \BibitemOpen
  \bibfield  {author} {\bibinfo {author} {\bibfnamefont {E.}~\bibnamefont
  {Kammann}}, \bibinfo {author} {\bibfnamefont {T.~C.~H.}\ \bibnamefont
  {Liew}}, \bibinfo {author} {\bibfnamefont {H.}~\bibnamefont {Ohadi}},
  \bibinfo {author} {\bibfnamefont {P.}~\bibnamefont {Cilibrizzi}}, \bibinfo
  {author} {\bibfnamefont {P.}~\bibnamefont {Tsotsis}}, \bibinfo {author}
  {\bibfnamefont {Z.}~\bibnamefont {Hatzopoulos}}, \bibinfo {author}
  {\bibfnamefont {P.~G.}\ \bibnamefont {Savvidis}}, \bibinfo {author}
  {\bibfnamefont {A.~V.}\ \bibnamefont {Kavokin}}, \ and\ \bibinfo {author}
  {\bibfnamefont {P.~G.}\ \bibnamefont {Lagoudakis}},\ }\href {\doibase
  10.1103/PhysRevLett.109.036404} {\bibfield  {journal} {\bibinfo  {journal}
  {Phys. Rev. Lett.}\ }\textbf {\bibinfo {volume} {109}},\ \bibinfo {pages}
  {036404} (\bibinfo {year} {2012})}\BibitemShut {NoStop}%
\bibitem [{\citenamefont {Dufferwiel}\ \emph {et~al.}(2015)\citenamefont
  {Dufferwiel}, \citenamefont {Li}, \citenamefont {Cancellieri}, \citenamefont
  {Giriunas}, \citenamefont {Trichet}, \citenamefont {Whittaker}, \citenamefont
  {Walker}, \citenamefont {Fras}, \citenamefont {Clarke}, \citenamefont
  {Smith}, \citenamefont {Skolnick},\ and\ \citenamefont
  {Krizhanovskii}}]{Dufferwiel2015}%
  \BibitemOpen
  \bibfield  {author} {\bibinfo {author} {\bibfnamefont {S.}~\bibnamefont
  {Dufferwiel}}, \bibinfo {author} {\bibfnamefont {F.}~\bibnamefont {Li}},
  \bibinfo {author} {\bibfnamefont {E.}~\bibnamefont {Cancellieri}}, \bibinfo
  {author} {\bibfnamefont {L.}~\bibnamefont {Giriunas}}, \bibinfo {author}
  {\bibfnamefont {A.~A.~P.}\ \bibnamefont {Trichet}}, \bibinfo {author}
  {\bibfnamefont {D.~M.}\ \bibnamefont {Whittaker}}, \bibinfo {author}
  {\bibfnamefont {P.~M.}\ \bibnamefont {Walker}}, \bibinfo {author}
  {\bibfnamefont {F.}~\bibnamefont {Fras}}, \bibinfo {author} {\bibfnamefont
  {E.}~\bibnamefont {Clarke}}, \bibinfo {author} {\bibfnamefont {J.~M.}\
  \bibnamefont {Smith}}, \bibinfo {author} {\bibfnamefont {M.~S.}\ \bibnamefont
  {Skolnick}}, \ and\ \bibinfo {author} {\bibfnamefont {D.~N.}\ \bibnamefont
  {Krizhanovskii}},\ }\href {\doibase 10.1103/PhysRevLett.115.246401}
  {\bibfield  {journal} {\bibinfo  {journal} {Phys. Rev. Lett.}\ }\textbf
  {\bibinfo {volume} {115}},\ \bibinfo {pages} {246401} (\bibinfo {year}
  {2015})}\BibitemShut {NoStop}%
\bibitem [{\citenamefont {Hivet}\ \emph {et~al.}(2012)\citenamefont {Hivet},
  \citenamefont {Flayac}, \citenamefont {Solnyshkov}, \citenamefont {Tanese},
  \citenamefont {Boulier}, \citenamefont {Andreoli}, \citenamefont {Giacobino},
  \citenamefont {Bloch}, \citenamefont {Bramati}, \citenamefont {Malpuech},\
  and\ \citenamefont {Amo}}]{hivet2012half}%
  \BibitemOpen
  \bibfield  {author} {\bibinfo {author} {\bibfnamefont {R.}~\bibnamefont
  {Hivet}}, \bibinfo {author} {\bibfnamefont {H.}~\bibnamefont {Flayac}},
  \bibinfo {author} {\bibfnamefont {D.~D.}\ \bibnamefont {Solnyshkov}},
  \bibinfo {author} {\bibfnamefont {D.}~\bibnamefont {Tanese}}, \bibinfo
  {author} {\bibfnamefont {T.}~\bibnamefont {Boulier}}, \bibinfo {author}
  {\bibfnamefont {D.}~\bibnamefont {Andreoli}}, \bibinfo {author}
  {\bibfnamefont {E.}~\bibnamefont {Giacobino}}, \bibinfo {author}
  {\bibfnamefont {J.}~\bibnamefont {Bloch}}, \bibinfo {author} {\bibfnamefont
  {A.}~\bibnamefont {Bramati}}, \bibinfo {author} {\bibfnamefont
  {G.}~\bibnamefont {Malpuech}}, \ and\ \bibinfo {author} {\bibfnamefont
  {A.}~\bibnamefont {Amo}},\ }\href {https://doi.org/10.1038/nphys2406}
  {\bibfield  {journal} {\bibinfo  {journal} {Nat. Phys.}\ }\textbf {\bibinfo
  {volume} {8}},\ \bibinfo {pages} {724} (\bibinfo {year} {2012})}\BibitemShut
  {NoStop}%
\bibitem [{\citenamefont {Liu}\ \emph {et~al.}(2015)\citenamefont {Liu},
  \citenamefont {Snoke}, \citenamefont {Daley}, \citenamefont {Pfeiffer},\ and\
  \citenamefont {West}}]{Gang2015}%
  \BibitemOpen
  \bibfield  {author} {\bibinfo {author} {\bibfnamefont {G.}~\bibnamefont
  {Liu}}, \bibinfo {author} {\bibfnamefont {D.~W.}\ \bibnamefont {Snoke}},
  \bibinfo {author} {\bibfnamefont {A.}~\bibnamefont {Daley}}, \bibinfo
  {author} {\bibfnamefont {L.~N.}\ \bibnamefont {Pfeiffer}}, \ and\ \bibinfo
  {author} {\bibfnamefont {K.}~\bibnamefont {West}},\ }\href {\doibase
  10.1073/pnas.1424549112} {\bibfield  {journal} {\bibinfo  {journal} {Proc.
  Natl. Acad. Sci.}\ }\textbf {\bibinfo {volume} {112}},\ \bibinfo {pages}
  {2676} (\bibinfo {year} {2015})}\BibitemShut {NoStop}%
\bibitem [{\citenamefont {Ohadi}\ \emph {et~al.}(2015)\citenamefont {Ohadi},
  \citenamefont {Dreismann}, \citenamefont {Rubo}, \citenamefont {Pinsker},
  \citenamefont {del Valle-Inclan~Redondo}, \citenamefont {Tsintzos},
  \citenamefont {Hatzopoulos}, \citenamefont {Savvidis},\ and\ \citenamefont
  {Baumberg}}]{Ohadi2015}%
  \BibitemOpen
  \bibfield  {author} {\bibinfo {author} {\bibfnamefont {H.}~\bibnamefont
  {Ohadi}}, \bibinfo {author} {\bibfnamefont {A.}~\bibnamefont {Dreismann}},
  \bibinfo {author} {\bibfnamefont {Y.~G.}\ \bibnamefont {Rubo}}, \bibinfo
  {author} {\bibfnamefont {F.}~\bibnamefont {Pinsker}}, \bibinfo {author}
  {\bibfnamefont {Y.}~\bibnamefont {del Valle-Inclan~Redondo}}, \bibinfo
  {author} {\bibfnamefont {S.~I.}\ \bibnamefont {Tsintzos}}, \bibinfo {author}
  {\bibfnamefont {Z.}~\bibnamefont {Hatzopoulos}}, \bibinfo {author}
  {\bibfnamefont {P.~G.}\ \bibnamefont {Savvidis}}, \ and\ \bibinfo {author}
  {\bibfnamefont {J.~J.}\ \bibnamefont {Baumberg}},\ }\href {\doibase
  10.1103/PhysRevX.5.031002} {\bibfield  {journal} {\bibinfo  {journal} {Phys.
  Rev. X}\ }\textbf {\bibinfo {volume} {5}},\ \bibinfo {pages} {031002}
  (\bibinfo {year} {2015})}\BibitemShut {NoStop}%
\bibitem [{\citenamefont {Baryshev}\ \emph {et~al.}(2022)\citenamefont
  {Baryshev}, \citenamefont {Zasedatelev}, \citenamefont {Sigurdsson},
  \citenamefont {Gnusov}, \citenamefont {T\"opfer}, \citenamefont
  {Askitopoulos},\ and\ \citenamefont {Lagoudakis}}]{Baryshev2022}%
  \BibitemOpen
  \bibfield  {author} {\bibinfo {author} {\bibfnamefont {S.}~\bibnamefont
  {Baryshev}}, \bibinfo {author} {\bibfnamefont {A.}~\bibnamefont
  {Zasedatelev}}, \bibinfo {author} {\bibfnamefont {H.}~\bibnamefont
  {Sigurdsson}}, \bibinfo {author} {\bibfnamefont {I.}~\bibnamefont {Gnusov}},
  \bibinfo {author} {\bibfnamefont {J.~D.}\ \bibnamefont {T\"opfer}}, \bibinfo
  {author} {\bibfnamefont {A.}~\bibnamefont {Askitopoulos}}, \ and\ \bibinfo
  {author} {\bibfnamefont {P.~G.}\ \bibnamefont {Lagoudakis}},\ }\href
  {\doibase 10.1103/PhysRevLett.128.087402} {\bibfield  {journal} {\bibinfo
  {journal} {Phys. Rev. Lett.}\ }\textbf {\bibinfo {volume} {128}},\ \bibinfo
  {pages} {087402} (\bibinfo {year} {2022})}\BibitemShut {NoStop}%
\bibitem [{\citenamefont {Chen}\ \emph {et~al.}(2018)\citenamefont {Chen},
  \citenamefont {Lin}, \citenamefont {Chen}, \citenamefont {Chiu},
  \citenamefont {Wang}, \citenamefont {Chen}, \citenamefont {Huang},
  \citenamefont {Yip}, \citenamefont {Kawaguchi},\ and\ \citenamefont
  {Lin}}]{Lin2018}%
  \BibitemOpen
  \bibfield  {author} {\bibinfo {author} {\bibfnamefont {H.-R.}\ \bibnamefont
  {Chen}}, \bibinfo {author} {\bibfnamefont {K.-Y.}\ \bibnamefont {Lin}},
  \bibinfo {author} {\bibfnamefont {P.-K.}\ \bibnamefont {Chen}}, \bibinfo
  {author} {\bibfnamefont {N.-C.}\ \bibnamefont {Chiu}}, \bibinfo {author}
  {\bibfnamefont {J.-B.}\ \bibnamefont {Wang}}, \bibinfo {author}
  {\bibfnamefont {C.-A.}\ \bibnamefont {Chen}}, \bibinfo {author}
  {\bibfnamefont {P.}~\bibnamefont {Huang}}, \bibinfo {author} {\bibfnamefont
  {S.-K.}\ \bibnamefont {Yip}}, \bibinfo {author} {\bibfnamefont
  {Y.}~\bibnamefont {Kawaguchi}}, \ and\ \bibinfo {author} {\bibfnamefont
  {Y.-J.}\ \bibnamefont {Lin}},\ }\href {\doibase
  10.1103/PhysRevLett.121.113204} {\bibfield  {journal} {\bibinfo  {journal}
  {Phys. Rev. Lett.}\ }\textbf {\bibinfo {volume} {121}},\ \bibinfo {pages}
  {113204} (\bibinfo {year} {2018})}\BibitemShut {NoStop}%
\bibitem [{\citenamefont {Zhang}\ \emph {et~al.}(2019)\citenamefont {Zhang},
  \citenamefont {Gao}, \citenamefont {Zou}, \citenamefont {Kong}, \citenamefont
  {Li}, \citenamefont {Shen}, \citenamefont {Chen}, \citenamefont {Peng},
  \citenamefont {Zhan}, \citenamefont {Pu},\ and\ \citenamefont
  {Jiang}}]{Jiang2019}%
  \BibitemOpen
  \bibfield  {author} {\bibinfo {author} {\bibfnamefont {D.}~\bibnamefont
  {Zhang}}, \bibinfo {author} {\bibfnamefont {T.}~\bibnamefont {Gao}}, \bibinfo
  {author} {\bibfnamefont {P.}~\bibnamefont {Zou}}, \bibinfo {author}
  {\bibfnamefont {L.}~\bibnamefont {Kong}}, \bibinfo {author} {\bibfnamefont
  {R.}~\bibnamefont {Li}}, \bibinfo {author} {\bibfnamefont {X.}~\bibnamefont
  {Shen}}, \bibinfo {author} {\bibfnamefont {X.-L.}\ \bibnamefont {Chen}},
  \bibinfo {author} {\bibfnamefont {S.-G.}\ \bibnamefont {Peng}}, \bibinfo
  {author} {\bibfnamefont {M.}~\bibnamefont {Zhan}}, \bibinfo {author}
  {\bibfnamefont {H.}~\bibnamefont {Pu}}, \ and\ \bibinfo {author}
  {\bibfnamefont {K.}~\bibnamefont {Jiang}},\ }\href {\doibase
  10.1103/PhysRevLett.122.110402} {\bibfield  {journal} {\bibinfo  {journal}
  {Phys. Rev. Lett.}\ }\textbf {\bibinfo {volume} {122}},\ \bibinfo {pages}
  {110402} (\bibinfo {year} {2019})}\BibitemShut {NoStop}%
\bibitem [{\citenamefont {Bidasyuk}\ \emph {et~al.}(2022)\citenamefont
  {Bidasyuk}, \citenamefont {Kovtunenko},\ and\ \citenamefont
  {Prikhodko}}]{Bidasyuk2022}%
  \BibitemOpen
  \bibfield  {author} {\bibinfo {author} {\bibfnamefont {Y.~M.}\ \bibnamefont
  {Bidasyuk}}, \bibinfo {author} {\bibfnamefont {K.~S.}\ \bibnamefont
  {Kovtunenko}}, \ and\ \bibinfo {author} {\bibfnamefont {O.~O.}\ \bibnamefont
  {Prikhodko}},\ }\href {\doibase 10.1103/PhysRevA.105.023320} {\bibfield
  {journal} {\bibinfo  {journal} {Phys. Rev. A}\ }\textbf {\bibinfo {volume}
  {105}},\ \bibinfo {pages} {023320} (\bibinfo {year} {2022})}\BibitemShut
  {NoStop}%
\bibitem [{\citenamefont {Wang}\ \emph {et~al.}(2021)\citenamefont {Wang},
  \citenamefont {Ji}, \citenamefont {Sun},\ and\ \citenamefont
  {Li}}]{Jian2021}%
  \BibitemOpen
  \bibfield  {author} {\bibinfo {author} {\bibfnamefont {L.-L.}\ \bibnamefont
  {Wang}}, \bibinfo {author} {\bibfnamefont {A.-C.}\ \bibnamefont {Ji}},
  \bibinfo {author} {\bibfnamefont {Q.}~\bibnamefont {Sun}}, \ and\ \bibinfo
  {author} {\bibfnamefont {J.}~\bibnamefont {Li}},\ }\href {\doibase
  10.1103/PhysRevLett.126.193401} {\bibfield  {journal} {\bibinfo  {journal}
  {Phys. Rev. Lett.}\ }\textbf {\bibinfo {volume} {126}},\ \bibinfo {pages}
  {193401} (\bibinfo {year} {2021})}\BibitemShut {NoStop}%
\bibitem [{\citenamefont {Chen}\ \emph
  {et~al.}(2020{\natexlab{a}})\citenamefont {Chen}, \citenamefont {Peng},
  \citenamefont {Zou}, \citenamefont {Liu},\ and\ \citenamefont
  {Hu}}]{Hui2020}%
  \BibitemOpen
  \bibfield  {author} {\bibinfo {author} {\bibfnamefont {X.-L.}\ \bibnamefont
  {Chen}}, \bibinfo {author} {\bibfnamefont {S.-G.}\ \bibnamefont {Peng}},
  \bibinfo {author} {\bibfnamefont {P.}~\bibnamefont {Zou}}, \bibinfo {author}
  {\bibfnamefont {X.-J.}\ \bibnamefont {Liu}}, \ and\ \bibinfo {author}
  {\bibfnamefont {H.}~\bibnamefont {Hu}},\ }\href {\doibase
  10.1103/PhysRevResearch.2.033152} {\bibfield  {journal} {\bibinfo  {journal}
  {Phys. Rev. Research}\ }\textbf {\bibinfo {volume} {2}},\ \bibinfo {pages}
  {033152} (\bibinfo {year} {2020}{\natexlab{a}})}\BibitemShut {NoStop}%
\bibitem [{\citenamefont {Chen}\ \emph
  {et~al.}(2020{\natexlab{b}})\citenamefont {Chen}, \citenamefont {Wu},
  \citenamefont {Peng}, \citenamefont {Yi},\ and\ \citenamefont
  {He}}]{Chen2020}%
  \BibitemOpen
  \bibfield  {author} {\bibinfo {author} {\bibfnamefont {K.-J.}\ \bibnamefont
  {Chen}}, \bibinfo {author} {\bibfnamefont {F.}~\bibnamefont {Wu}}, \bibinfo
  {author} {\bibfnamefont {S.-G.}\ \bibnamefont {Peng}}, \bibinfo {author}
  {\bibfnamefont {W.}~\bibnamefont {Yi}}, \ and\ \bibinfo {author}
  {\bibfnamefont {L.}~\bibnamefont {He}},\ }\href {\doibase
  10.1103/PhysRevLett.125.260407} {\bibfield  {journal} {\bibinfo  {journal}
  {Phys. Rev. Lett.}\ }\textbf {\bibinfo {volume} {125}},\ \bibinfo {pages}
  {260407} (\bibinfo {year} {2020}{\natexlab{b}})}\BibitemShut {NoStop}%
\bibitem [{\citenamefont {Bruchhausen}\ \emph {et~al.}(2008)\citenamefont
  {Bruchhausen}, \citenamefont {Le\'{o}n~Hilario}, \citenamefont {Aligia},
  \citenamefont {Lobos}, \citenamefont {Fainstein}, \citenamefont {Jusserand},\
  and\ \citenamefont {Andr\'e}}]{Bruchhausen2008}%
  \BibitemOpen
  \bibfield  {author} {\bibinfo {author} {\bibfnamefont {A.}~\bibnamefont
  {Bruchhausen}}, \bibinfo {author} {\bibfnamefont {L.~M.}\ \bibnamefont
  {Le\'{o}n~Hilario}}, \bibinfo {author} {\bibfnamefont {A.~A.}\ \bibnamefont
  {Aligia}}, \bibinfo {author} {\bibfnamefont {A.~M.}\ \bibnamefont {Lobos}},
  \bibinfo {author} {\bibfnamefont {A.}~\bibnamefont {Fainstein}}, \bibinfo
  {author} {\bibfnamefont {B.}~\bibnamefont {Jusserand}}, \ and\ \bibinfo
  {author} {\bibfnamefont {R.}~\bibnamefont {Andr\'e}},\ }\href {\doibase
  10.1103/PhysRevB.78.125326} {\bibfield  {journal} {\bibinfo  {journal} {Phys.
  Rev. B}\ }\textbf {\bibinfo {volume} {78}},\ \bibinfo {pages} {125326}
  (\bibinfo {year} {2008})}\BibitemShut {NoStop}%
\bibitem [{\citenamefont {Yang}\ and\ \citenamefont {Kim}(2022)}]{Yang2022}%
  \BibitemOpen
  \bibfield  {author} {\bibinfo {author} {\bibfnamefont {H.}~\bibnamefont
  {Yang}}\ and\ \bibinfo {author} {\bibfnamefont {N.~Y.}\ \bibnamefont {Kim}},\
  }\href {\doibase https://doi.org/10.1002/qute.202100137} {\bibfield
  {journal} {\bibinfo  {journal} {Adv. Quantum Technol.}\ }\textbf {\bibinfo
  {volume} {5}},\ \bibinfo {pages} {2100137} (\bibinfo {year}
  {2022})}\BibitemShut {NoStop}%
\bibitem [{\citenamefont {Glazov}\ \emph {et~al.}(2015)\citenamefont {Glazov},
  \citenamefont {Ivchenko}, \citenamefont {Wang}, \citenamefont {Amand},
  \citenamefont {Marie}, \citenamefont {Urbaszek},\ and\ \citenamefont
  {Liu}}]{Glazov2015}%
  \BibitemOpen
  \bibfield  {author} {\bibinfo {author} {\bibfnamefont {M.~M.}\ \bibnamefont
  {Glazov}}, \bibinfo {author} {\bibfnamefont {E.~L.}\ \bibnamefont
  {Ivchenko}}, \bibinfo {author} {\bibfnamefont {G.}~\bibnamefont {Wang}},
  \bibinfo {author} {\bibfnamefont {T.}~\bibnamefont {Amand}}, \bibinfo
  {author} {\bibfnamefont {X.}~\bibnamefont {Marie}}, \bibinfo {author}
  {\bibfnamefont {B.}~\bibnamefont {Urbaszek}}, \ and\ \bibinfo {author}
  {\bibfnamefont {B.~L.}\ \bibnamefont {Liu}},\ }\href {\doibase
  https://doi.org/10.1002/pssb.201552211} {\bibfield  {journal} {\bibinfo
  {journal} {Phys. Status. Solidi. (b)}\ }\textbf {\bibinfo {volume} {252}},\
  \bibinfo {pages} {2349} (\bibinfo {year} {2015})}\BibitemShut {NoStop}%
\bibitem [{\citenamefont {Maialle}\ \emph {et~al.}(1993)\citenamefont
  {Maialle}, \citenamefont {de~Andrada~e Silva},\ and\ \citenamefont
  {Sham}}]{Maialle1993}%
  \BibitemOpen
  \bibfield  {author} {\bibinfo {author} {\bibfnamefont {M.~Z.}\ \bibnamefont
  {Maialle}}, \bibinfo {author} {\bibfnamefont {E.~A.}\ \bibnamefont
  {de~Andrada~e Silva}}, \ and\ \bibinfo {author} {\bibfnamefont {L.~J.}\
  \bibnamefont {Sham}},\ }\href {\doibase 10.1103/PhysRevB.47.15776} {\bibfield
   {journal} {\bibinfo  {journal} {Phys. Rev. B}\ }\textbf {\bibinfo {volume}
  {47}},\ \bibinfo {pages} {15776} (\bibinfo {year} {1993})}\BibitemShut
  {NoStop}%
\bibitem [{\citenamefont {Baxter}\ \emph {et~al.}(1997)\citenamefont {Baxter},
  \citenamefont {Skolnick}, \citenamefont {Armitage}, \citenamefont {Astratov},
  \citenamefont {Whittaker}, \citenamefont {Fisher}, \citenamefont {Roberts},
  \citenamefont {Mowbray},\ and\ \citenamefont {Kaliteevski}}]{Baxter1997}%
  \BibitemOpen
  \bibfield  {author} {\bibinfo {author} {\bibfnamefont {D.}~\bibnamefont
  {Baxter}}, \bibinfo {author} {\bibfnamefont {M.~S.}\ \bibnamefont
  {Skolnick}}, \bibinfo {author} {\bibfnamefont {A.}~\bibnamefont {Armitage}},
  \bibinfo {author} {\bibfnamefont {V.~N.}\ \bibnamefont {Astratov}}, \bibinfo
  {author} {\bibfnamefont {D.~M.}\ \bibnamefont {Whittaker}}, \bibinfo {author}
  {\bibfnamefont {T.~A.}\ \bibnamefont {Fisher}}, \bibinfo {author}
  {\bibfnamefont {J.~S.}\ \bibnamefont {Roberts}}, \bibinfo {author}
  {\bibfnamefont {D.~J.}\ \bibnamefont {Mowbray}}, \ and\ \bibinfo {author}
  {\bibfnamefont {M.~A.}\ \bibnamefont {Kaliteevski}},\ }\href {\doibase
  10.1103/PhysRevB.56.R10032} {\bibfield  {journal} {\bibinfo  {journal} {Phys.
  Rev. B}\ }\textbf {\bibinfo {volume} {56}},\ \bibinfo {pages} {R10032}
  (\bibinfo {year} {1997})}\BibitemShut {NoStop}%
\bibitem [{\citenamefont {Levrat}\ \emph {et~al.}(2010)\citenamefont {Levrat},
  \citenamefont {Butt\'e}, \citenamefont {Christian}, \citenamefont {Glauser},
  \citenamefont {Feltin}, \citenamefont {Carlin}, \citenamefont {Grandjean},
  \citenamefont {Read}, \citenamefont {Kavokin},\ and\ \citenamefont
  {Rubo}}]{Levrat2010}%
  \BibitemOpen
  \bibfield  {author} {\bibinfo {author} {\bibfnamefont {J.}~\bibnamefont
  {Levrat}}, \bibinfo {author} {\bibfnamefont {R.}~\bibnamefont {Butt\'e}},
  \bibinfo {author} {\bibfnamefont {T.}~\bibnamefont {Christian}}, \bibinfo
  {author} {\bibfnamefont {M.}~\bibnamefont {Glauser}}, \bibinfo {author}
  {\bibfnamefont {E.}~\bibnamefont {Feltin}}, \bibinfo {author} {\bibfnamefont
  {J.-F.}\ \bibnamefont {Carlin}}, \bibinfo {author} {\bibfnamefont
  {N.}~\bibnamefont {Grandjean}}, \bibinfo {author} {\bibfnamefont
  {D.}~\bibnamefont {Read}}, \bibinfo {author} {\bibfnamefont {A.~V.}\
  \bibnamefont {Kavokin}}, \ and\ \bibinfo {author} {\bibfnamefont {Y.~G.}\
  \bibnamefont {Rubo}},\ }\href {\doibase 10.1103/PhysRevLett.104.166402}
  {\bibfield  {journal} {\bibinfo  {journal} {Phys. Rev. Lett.}\ }\textbf
  {\bibinfo {volume} {104}},\ \bibinfo {pages} {166402} (\bibinfo {year}
  {2010})}\BibitemShut {NoStop}%
\bibitem [{\citenamefont {Pinsker}\ and\ \citenamefont
  {Flayac}(2014)}]{Pinsker2014}%
  \BibitemOpen
  \bibfield  {author} {\bibinfo {author} {\bibfnamefont {F.}~\bibnamefont
  {Pinsker}}\ and\ \bibinfo {author} {\bibfnamefont {H.}~\bibnamefont
  {Flayac}},\ }\href {\doibase 10.1103/PhysRevLett.112.140405} {\bibfield
  {journal} {\bibinfo  {journal} {Phys. Rev. Lett.}\ }\textbf {\bibinfo
  {volume} {112}},\ \bibinfo {pages} {140405} (\bibinfo {year}
  {2014})}\BibitemShut {NoStop}%
\bibitem [{\citenamefont {Li}\ \emph {et~al.}(2015)\citenamefont {Li},
  \citenamefont {Liew}, \citenamefont {Egorov},\ and\ \citenamefont
  {Ostrovskaya}}]{Li2015}%
  \BibitemOpen
  \bibfield  {author} {\bibinfo {author} {\bibfnamefont {G.}~\bibnamefont
  {Li}}, \bibinfo {author} {\bibfnamefont {T.~C.~H.}\ \bibnamefont {Liew}},
  \bibinfo {author} {\bibfnamefont {O.~A.}\ \bibnamefont {Egorov}}, \ and\
  \bibinfo {author} {\bibfnamefont {E.~A.}\ \bibnamefont {Ostrovskaya}},\
  }\href {\doibase 10.1103/PhysRevB.92.064304} {\bibfield  {journal} {\bibinfo
  {journal} {Phys. Rev. B}\ }\textbf {\bibinfo {volume} {92}},\ \bibinfo
  {pages} {064304} (\bibinfo {year} {2015})}\BibitemShut {NoStop}%
\bibitem [{\citenamefont {Yulin}\ \emph {et~al.}(2016)\citenamefont {Yulin},
  \citenamefont {Desyatnikov},\ and\ \citenamefont {Ostrovskaya}}]{Yulin2016}%
  \BibitemOpen
  \bibfield  {author} {\bibinfo {author} {\bibfnamefont {A.~V.}\ \bibnamefont
  {Yulin}}, \bibinfo {author} {\bibfnamefont {A.~S.}\ \bibnamefont
  {Desyatnikov}}, \ and\ \bibinfo {author} {\bibfnamefont {E.~A.}\ \bibnamefont
  {Ostrovskaya}},\ }\href {\doibase 10.1103/PhysRevB.94.134310} {\bibfield
  {journal} {\bibinfo  {journal} {Phys. Rev. B}\ }\textbf {\bibinfo {volume}
  {94}},\ \bibinfo {pages} {134310} (\bibinfo {year} {2016})}\BibitemShut
  {NoStop}%
\bibitem [{\citenamefont {Carusotto}\ and\ \citenamefont
  {Ciuti}(2013)}]{RevModPhys.85.299}%
  \BibitemOpen
  \bibfield  {author} {\bibinfo {author} {\bibfnamefont {I.}~\bibnamefont
  {Carusotto}}\ and\ \bibinfo {author} {\bibfnamefont {C.}~\bibnamefont
  {Ciuti}},\ }\href {\doibase 10.1103/RevModPhys.85.299} {\bibfield  {journal}
  {\bibinfo  {journal} {Rev. Mod. Phys.}\ }\textbf {\bibinfo {volume} {85}},\
  \bibinfo {pages} {299} (\bibinfo {year} {2013})}\BibitemShut {NoStop}%
\bibitem [{\citenamefont {Dum}\ \emph {et~al.}(1998)\citenamefont {Dum},
  \citenamefont {Cirac}, \citenamefont {Lewenstein},\ and\ \citenamefont
  {Zoller}}]{Dum1998}%
  \BibitemOpen
  \bibfield  {author} {\bibinfo {author} {\bibfnamefont {R.}~\bibnamefont
  {Dum}}, \bibinfo {author} {\bibfnamefont {J.~I.}\ \bibnamefont {Cirac}},
  \bibinfo {author} {\bibfnamefont {M.}~\bibnamefont {Lewenstein}}, \ and\
  \bibinfo {author} {\bibfnamefont {P.}~\bibnamefont {Zoller}},\ }\href
  {\doibase 10.1103/PhysRevLett.80.2972} {\bibfield  {journal} {\bibinfo
  {journal} {Phys. Rev. Lett.}\ }\textbf {\bibinfo {volume} {80}},\ \bibinfo
  {pages} {2972} (\bibinfo {year} {1998})}\BibitemShut {NoStop}%
\bibitem [{\citenamefont {Marzlin}\ \emph {et~al.}(1997)\citenamefont
  {Marzlin}, \citenamefont {Zhang},\ and\ \citenamefont
  {Wright}}]{Marzlin1997}%
  \BibitemOpen
  \bibfield  {author} {\bibinfo {author} {\bibfnamefont {K.-P.}\ \bibnamefont
  {Marzlin}}, \bibinfo {author} {\bibfnamefont {W.}~\bibnamefont {Zhang}}, \
  and\ \bibinfo {author} {\bibfnamefont {E.~M.}\ \bibnamefont {Wright}},\
  }\href {\doibase 10.1103/PhysRevLett.79.4728} {\bibfield  {journal} {\bibinfo
   {journal} {Phys. Rev. Lett.}\ }\textbf {\bibinfo {volume} {79}},\ \bibinfo
  {pages} {4728} (\bibinfo {year} {1997})}\BibitemShut {NoStop}%
\bibitem [{\citenamefont {Dutton}\ and\ \citenamefont
  {Ruostekoski}(2004)}]{Dutton2004}%
  \BibitemOpen
  \bibfield  {author} {\bibinfo {author} {\bibfnamefont {Z.}~\bibnamefont
  {Dutton}}\ and\ \bibinfo {author} {\bibfnamefont {J.}~\bibnamefont
  {Ruostekoski}},\ }\href {\doibase 10.1103/PhysRevLett.93.193602} {\bibfield
  {journal} {\bibinfo  {journal} {Phys. Rev. Lett.}\ }\textbf {\bibinfo
  {volume} {93}},\ \bibinfo {pages} {193602} (\bibinfo {year}
  {2004})}\BibitemShut {NoStop}%
\bibitem [{\citenamefont {Maucher}\ \emph {et~al.}(2018)\citenamefont
  {Maucher}, \citenamefont {Skupin}, \citenamefont {Gardiner},\ and\
  \citenamefont {Hughes}}]{Maucher2018}%
  \BibitemOpen
  \bibfield  {author} {\bibinfo {author} {\bibfnamefont {F.}~\bibnamefont
  {Maucher}}, \bibinfo {author} {\bibfnamefont {S.}~\bibnamefont {Skupin}},
  \bibinfo {author} {\bibfnamefont {S.~A.}\ \bibnamefont {Gardiner}}, \ and\
  \bibinfo {author} {\bibfnamefont {I.~G.}\ \bibnamefont {Hughes}},\ }\href
  {\doibase 10.1103/PhysRevLett.120.163903} {\bibfield  {journal} {\bibinfo
  {journal} {Phys. Rev. Lett.}\ }\textbf {\bibinfo {volume} {120}},\ \bibinfo
  {pages} {163903} (\bibinfo {year} {2018})}\BibitemShut {NoStop}%
\bibitem [{\citenamefont {Xu}\ \emph {et~al.}(2017)\citenamefont {Xu},
  \citenamefont {Hu}, \citenamefont {Zhang},\ and\ \citenamefont
  {Liang}}]{Liang2017}%
  \BibitemOpen
  \bibfield  {author} {\bibinfo {author} {\bibfnamefont {X.}~\bibnamefont
  {Xu}}, \bibinfo {author} {\bibfnamefont {Y.}~\bibnamefont {Hu}}, \bibinfo
  {author} {\bibfnamefont {Z.}~\bibnamefont {Zhang}}, \ and\ \bibinfo {author}
  {\bibfnamefont {Z.}~\bibnamefont {Liang}},\ }\href {\doibase
  10.1103/PhysRevB.96.144511} {\bibfield  {journal} {\bibinfo  {journal} {Phys.
  Rev. B}\ }\textbf {\bibinfo {volume} {96}},\ \bibinfo {pages} {144511}
  (\bibinfo {year} {2017})}\BibitemShut {NoStop}%
\bibitem [{\citenamefont {Borgh}\ \emph {et~al.}(2010)\citenamefont {Borgh},
  \citenamefont {Keeling},\ and\ \citenamefont {Berloff}}]{Borgh2010}%
  \BibitemOpen
  \bibfield  {author} {\bibinfo {author} {\bibfnamefont {M.~O.}\ \bibnamefont
  {Borgh}}, \bibinfo {author} {\bibfnamefont {J.}~\bibnamefont {Keeling}}, \
  and\ \bibinfo {author} {\bibfnamefont {N.~G.}\ \bibnamefont {Berloff}},\
  }\href {\doibase 10.1103/PhysRevB.81.235302} {\bibfield  {journal} {\bibinfo
  {journal} {Phys. Rev. B}\ }\textbf {\bibinfo {volume} {81}},\ \bibinfo
  {pages} {235302} (\bibinfo {year} {2010})}\BibitemShut {NoStop}%
\bibitem [{\citenamefont {Sturm}\ \emph {et~al.}(2015)\citenamefont {Sturm},
  \citenamefont {Solnyshkov}, \citenamefont {Krebs}, \citenamefont
  {Lema\^{\i}tre}, \citenamefont {Sagnes}, \citenamefont {Galopin},
  \citenamefont {Amo}, \citenamefont {Malpuech},\ and\ \citenamefont
  {Bloch}}]{Sturm2015}%
  \BibitemOpen
  \bibfield  {author} {\bibinfo {author} {\bibfnamefont {C.}~\bibnamefont
  {Sturm}}, \bibinfo {author} {\bibfnamefont {D.}~\bibnamefont {Solnyshkov}},
  \bibinfo {author} {\bibfnamefont {O.}~\bibnamefont {Krebs}}, \bibinfo
  {author} {\bibfnamefont {A.}~\bibnamefont {Lema\^{\i}tre}}, \bibinfo {author}
  {\bibfnamefont {I.}~\bibnamefont {Sagnes}}, \bibinfo {author} {\bibfnamefont
  {E.}~\bibnamefont {Galopin}}, \bibinfo {author} {\bibfnamefont
  {A.}~\bibnamefont {Amo}}, \bibinfo {author} {\bibfnamefont {G.}~\bibnamefont
  {Malpuech}}, \ and\ \bibinfo {author} {\bibfnamefont {J.}~\bibnamefont
  {Bloch}},\ }\href {\doibase 10.1103/PhysRevB.91.155130} {\bibfield  {journal}
  {\bibinfo  {journal} {Phys. Rev. B}\ }\textbf {\bibinfo {volume} {91}},\
  \bibinfo {pages} {155130} (\bibinfo {year} {2015})}\BibitemShut {NoStop}%
\bibitem [{\citenamefont {Balili}\ \emph {et~al.}(2007)\citenamefont {Balili},
  \citenamefont {Hartwell}, \citenamefont {Snoke}, \citenamefont {Pfeiffer},\
  and\ \citenamefont {West}}]{Science2007}%
  \BibitemOpen
  \bibfield  {author} {\bibinfo {author} {\bibfnamefont {R.}~\bibnamefont
  {Balili}}, \bibinfo {author} {\bibfnamefont {V.}~\bibnamefont {Hartwell}},
  \bibinfo {author} {\bibfnamefont {D.}~\bibnamefont {Snoke}}, \bibinfo
  {author} {\bibfnamefont {L.}~\bibnamefont {Pfeiffer}}, \ and\ \bibinfo
  {author} {\bibfnamefont {K.}~\bibnamefont {West}},\ }\href {\doibase
  10.1126/science.1140990} {\bibfield  {journal} {\bibinfo  {journal}
  {Science}\ }\textbf {\bibinfo {volume} {316}},\ \bibinfo {pages} {1007}
  (\bibinfo {year} {2007})}\BibitemShut {NoStop}%
\bibitem [{\citenamefont {Keeling}\ and\ \citenamefont
  {Berloff}(2008)}]{Keeling2008}%
  \BibitemOpen
  \bibfield  {author} {\bibinfo {author} {\bibfnamefont {J.}~\bibnamefont
  {Keeling}}\ and\ \bibinfo {author} {\bibfnamefont {N.~G.}\ \bibnamefont
  {Berloff}},\ }\href {\doibase 10.1103/PhysRevLett.100.250401} {\bibfield
  {journal} {\bibinfo  {journal} {Phys. Rev. Lett.}\ }\textbf {\bibinfo
  {volume} {100}},\ \bibinfo {pages} {250401} (\bibinfo {year}
  {2008})}\BibitemShut {NoStop}%
\bibitem [{\citenamefont {Zhai}\ \emph {et~al.}(2023)\citenamefont {Zhai},
  \citenamefont {Ma}, \citenamefont {Gao}, \citenamefont {Xing}, \citenamefont
  {Gao}, \citenamefont {Dai}, \citenamefont {Wang}, \citenamefont {Pan},
  \citenamefont {Schumacher},\ and\ \citenamefont
  {Gao}}]{PhysRevLett.131.136901}%
  \BibitemOpen
  \bibfield  {author} {\bibinfo {author} {\bibfnamefont {X.}~\bibnamefont
  {Zhai}}, \bibinfo {author} {\bibfnamefont {X.}~\bibnamefont {Ma}}, \bibinfo
  {author} {\bibfnamefont {Y.}~\bibnamefont {Gao}}, \bibinfo {author}
  {\bibfnamefont {C.}~\bibnamefont {Xing}}, \bibinfo {author} {\bibfnamefont
  {M.}~\bibnamefont {Gao}}, \bibinfo {author} {\bibfnamefont {H.}~\bibnamefont
  {Dai}}, \bibinfo {author} {\bibfnamefont {X.}~\bibnamefont {Wang}}, \bibinfo
  {author} {\bibfnamefont {A.}~\bibnamefont {Pan}}, \bibinfo {author}
  {\bibfnamefont {S.}~\bibnamefont {Schumacher}}, \ and\ \bibinfo {author}
  {\bibfnamefont {T.}~\bibnamefont {Gao}},\ }\href {\doibase
  10.1103/PhysRevLett.131.136901} {\bibfield  {journal} {\bibinfo  {journal}
  {Phys. Rev. Lett.}\ }\textbf {\bibinfo {volume} {131}},\ \bibinfo {pages}
  {136901} (\bibinfo {year} {2023})}\BibitemShut {NoStop}%
\bibitem [{\citenamefont {Roumpos}\ \emph {et~al.}(2011)\citenamefont
  {Roumpos}, \citenamefont {Fraser}, \citenamefont {L\"{o}ffler}, \citenamefont
  {H\"{o}fling}, \citenamefont {Forchel},\ and\ \citenamefont
  {Yamamoto}}]{Georgios2011}%
  \BibitemOpen
  \bibfield  {author} {\bibinfo {author} {\bibfnamefont {G.}~\bibnamefont
  {Roumpos}}, \bibinfo {author} {\bibfnamefont {M.~D.}\ \bibnamefont {Fraser}},
  \bibinfo {author} {\bibfnamefont {A.}~\bibnamefont {L\"{o}ffler}}, \bibinfo
  {author} {\bibfnamefont {S.}~\bibnamefont {H\"{o}fling}}, \bibinfo {author}
  {\bibfnamefont {A.}~\bibnamefont {Forchel}}, \ and\ \bibinfo {author}
  {\bibfnamefont {Y.}~\bibnamefont {Yamamoto}},\ }\href
  {https://doi.org/10.1038/nphys1841} {\bibfield  {journal} {\bibinfo
  {journal} {Nature Phys.}\ }\textbf {\bibinfo {volume} {7}},\ \bibinfo {pages}
  {129} (\bibinfo {year} {2011})}\BibitemShut {NoStop}%
\bibitem [{\citenamefont {Borgh}\ \emph {et~al.}(2012)\citenamefont {Borgh},
  \citenamefont {Franchetti}, \citenamefont {Keeling},\ and\ \citenamefont
  {Berloff}}]{Borgh2012}%
  \BibitemOpen
  \bibfield  {author} {\bibinfo {author} {\bibfnamefont {M.~O.}\ \bibnamefont
  {Borgh}}, \bibinfo {author} {\bibfnamefont {G.}~\bibnamefont {Franchetti}},
  \bibinfo {author} {\bibfnamefont {J.}~\bibnamefont {Keeling}}, \ and\
  \bibinfo {author} {\bibfnamefont {N.~G.}\ \bibnamefont {Berloff}},\ }\href
  {\doibase 10.1103/PhysRevB.86.035307} {\bibfield  {journal} {\bibinfo
  {journal} {Phys. Rev. B}\ }\textbf {\bibinfo {volume} {86}},\ \bibinfo
  {pages} {035307} (\bibinfo {year} {2012})}\BibitemShut {NoStop}%
\bibitem [{\citenamefont {Bao}\ and\ \citenamefont
  {Wang}(2006)}]{bao2006efficient}%
  \BibitemOpen
  \bibfield  {author} {\bibinfo {author} {\bibfnamefont {W.}~\bibnamefont
  {Bao}}\ and\ \bibinfo {author} {\bibfnamefont {H.}~\bibnamefont {Wang}},\
  }\href {\doibase https://doi.org/10.1016/j.jcp.2006.01.020} {\bibfield
  {journal} {\bibinfo  {journal} {J. Comput. Phys.}\ }\textbf {\bibinfo
  {volume} {217}},\ \bibinfo {pages} {612} (\bibinfo {year}
  {2006})}\BibitemShut {NoStop}%
\bibitem [{\citenamefont {Fetter}(2009)}]{Fetter2009}%
  \BibitemOpen
  \bibfield  {author} {\bibinfo {author} {\bibfnamefont {A.~L.}\ \bibnamefont
  {Fetter}},\ }\href {\doibase 10.1103/RevModPhys.81.647} {\bibfield  {journal}
  {\bibinfo  {journal} {Rev. Mod. Phys.}\ }\textbf {\bibinfo {volume} {81}},\
  \bibinfo {pages} {647} (\bibinfo {year} {2009})}\BibitemShut {NoStop}%
\bibitem [{\citenamefont {Rooney}\ \emph {et~al.}(2010)\citenamefont {Rooney},
  \citenamefont {Bradley},\ and\ \citenamefont {Blakie}}]{Rooney2010}%
  \BibitemOpen
  \bibfield  {author} {\bibinfo {author} {\bibfnamefont {S.~J.}\ \bibnamefont
  {Rooney}}, \bibinfo {author} {\bibfnamefont {A.~S.}\ \bibnamefont {Bradley}},
  \ and\ \bibinfo {author} {\bibfnamefont {P.~B.}\ \bibnamefont {Blakie}},\
  }\href {\doibase 10.1103/PhysRevA.81.023630} {\bibfield  {journal} {\bibinfo
  {journal} {Phys. Rev. A}\ }\textbf {\bibinfo {volume} {81}},\ \bibinfo
  {pages} {023630} (\bibinfo {year} {2010})}\BibitemShut {NoStop}%
\bibitem [{\citenamefont {Sedov}\ \emph {et~al.}(2021)\citenamefont {Sedov},
  \citenamefont {Lukoshkin}, \citenamefont {Kalevich}, \citenamefont
  {Savvidis},\ and\ \citenamefont {Kavokin}}]{Sedov2021}%
  \BibitemOpen
  \bibfield  {author} {\bibinfo {author} {\bibfnamefont {E.~S.}\ \bibnamefont
  {Sedov}}, \bibinfo {author} {\bibfnamefont {V.~A.}\ \bibnamefont
  {Lukoshkin}}, \bibinfo {author} {\bibfnamefont {V.~K.}\ \bibnamefont
  {Kalevich}}, \bibinfo {author} {\bibfnamefont {P.~G.}\ \bibnamefont
  {Savvidis}}, \ and\ \bibinfo {author} {\bibfnamefont {A.~V.}\ \bibnamefont
  {Kavokin}},\ }\href {\doibase 10.1103/PhysRevResearch.3.013072} {\bibfield
  {journal} {\bibinfo  {journal} {Phys. Rev. Research}\ }\textbf {\bibinfo
  {volume} {3}},\ \bibinfo {pages} {013072} (\bibinfo {year}
  {2021})}\BibitemShut {NoStop}%
\bibitem [{\citenamefont {Ma}\ \emph {et~al.}(2020)\citenamefont {Ma},
  \citenamefont {Kartashov}, \citenamefont {Gao}, \citenamefont {Torner},\ and\
  \citenamefont {Schumacher}}]{Xuekai2020}%
  \BibitemOpen
  \bibfield  {author} {\bibinfo {author} {\bibfnamefont {X.}~\bibnamefont
  {Ma}}, \bibinfo {author} {\bibfnamefont {Y.~V.}\ \bibnamefont {Kartashov}},
  \bibinfo {author} {\bibfnamefont {T.}~\bibnamefont {Gao}}, \bibinfo {author}
  {\bibfnamefont {L.}~\bibnamefont {Torner}}, \ and\ \bibinfo {author}
  {\bibfnamefont {S.}~\bibnamefont {Schumacher}},\ }\href {\doibase
  10.1103/PhysRevB.102.045309} {\bibfield  {journal} {\bibinfo  {journal}
  {Phys. Rev. B}\ }\textbf {\bibinfo {volume} {102}},\ \bibinfo {pages}
  {045309} (\bibinfo {year} {2020})}\BibitemShut {NoStop}%
\end{thebibliography}%
%

\end{document}